\documentclass[aps,prd,notitlepage,showpacs,nofootinbib,superscriptaddress]{revtex4-1}
\pdfoutput=1
\usepackage{graphicx}
\usepackage{tabularx}
\usepackage[utf8]{inputenc} 
\usepackage{amsmath}
\usepackage{amssymb}
\usepackage{float}
\usepackage{comment}
\usepackage{slashed}
\usepackage[normalem]{ulem}
\usepackage{caption}
\usepackage{subcaption}
\usepackage[usenames,dvipsnames]{color}
\usepackage[colorinlistoftodos]{todonotes}
\usepackage[colorlinks=true,citecolor=darkred,urlcolor=darkred, pdfborder={0 0 0}]{hyperref}
\usepackage[normalem]{ulem}
\usepackage[colorinlistoftodos]{todonotes}

\newcommand {\ignore}[1]{}

\usepackage{epsf}
\usepackage{epsfig}
\usepackage{multirow}
\usepackage{subeqnarray}
\def\met {{\not\!\! E_T}}
\usepackage{scalefnt}

\definecolor{mightnightblue}{RGB}{25,25,112}
\definecolor{brown}{rgb}{0.59, 0.29, 0.0}
\definecolor{nicered}{rgb}{0.8,0.1,0.1}
\definecolor{nicegreen}{rgb}{0.1,0.5,0.1}
\definecolor{linkcolor}{rgb}{0,0,0.5}
\definecolor{darkred}{rgb}{0.6,0,0}

\def\gsim{\raise0.3ex\hbox{$\;>$\kern-0.75em\raise-1.1ex\hbox{$\sim\;$}}}
\def\lsim{\raise0.3ex\hbox{$\;<$\kern-0.75em\raise-1.1ex\hbox{$\sim\;$}}}
\newcommand{\sm}{{Standard Model }}

\usepackage{soul}

\newcommand {\black} {\color{black}}

\def\lfv{lepton flavour violation }

\def\vev#1{\left\langle #1\right\rangle}

\def\21{$\mathrm{SU(2)_L \otimes U(1)_Y}$}
\def\lfv{lepton flavor violation }

\def\sm{standard model }

\def\cos{\rm {cos}}
\def\sin{\rm {sin}}

\definecolor{vdrgreen}{rgb}{0.0, 0.7, 0.0}

\newcommand{\AddrAHEP}{%
  AHEP Group, Institut de F\'{i}sica Corpuscular --
  CSIC/Universitat de Val\`{e}ncia, Parc Cient\'ific de Paterna.\\
 C/ Catedr\'atico Jos\'e Beltr\'an, 2 E-46980 Paterna (Valencia) - Spain}

\begin{document}


\title{\boldmath \color{BrickRed} Phenomenology of scotogenic scalar dark matter}

\author{Ivania M. \'Avila}
\email{idmaturana@uc.cl}
\affiliation{Instituto de F\'{\i}sica, Pontificia Universidad Cat\'olica de Chile, Av. Vicuña Mackenna 4860, Santiago de Chile - Chile}
\affiliation{\AddrAHEP}

\author{Valentina De Romeri}
\email{deromeri@ific.uv.es}
\affiliation{\AddrAHEP}

\author{Laura Duarte}
\email{l.duarte@unesp.br}
\affiliation{Departamento de F\'isica e Qu\'imica, Universidade Estadual Paulista (UNESP),\\Guaratinguet\'a-SP - Brazil}
\affiliation{\AddrAHEP}

\author{Jos\'{e} W. F. Valle}\email{valle@ific.uv.es}
\affiliation{\AddrAHEP}

\begin{abstract}
  \vspace{0.5cm} We reexamine the minimal Singlet + Triplet Scotogenic Model, where dark matter is the mediator of neutrino mass generation.
    We assume it to be a scalar WIMP, whose stability follows from the same $\mathbb{Z} _{2}$ symmetry that leads to the radiative origin of neutrino masses.
    The scheme is the minimal one that allows for solar and atmospheric mass scales to be generated.
    We perform a full numerical analysis of the signatures expected at dark matter as well as collider experiments.
    We identify parameter regions where dark matter predictions agree with theoretical and experimental constraints, such as neutrino oscillations, Higgs data, dark matter relic
    abundance and direct detection searches. We also present forecasts for near future direct and indirect detection experiments. 
    These will further probe the parameter space. Finally, we explore collider signatures associated with the mono-jet channel at the LHC,
highlighting the existence of a viable light dark matter mass range.

  \end{abstract}

\maketitle
\noindent

\section{Introduction}
\label{Sect:intro}

After the discovery of the Higgs boson, the particle physics community is eager to discover new phenomena, that would imply physics beyond the Standard Model (SM).
Together with the evidence for dark matter, neutrino physics remains as the most solid indication of new physics.
Neutrino experiments point towards two different neutrino mass squared differences, associated to solar and atmospheric oscillations.
Hence, at least two of the three active neutrino species must be massive. Here we adopt the minimal picture in which one of the neutrinos is (nearly) massless.
This is achieved in ``missing partner'' seesaw mechanisms~\cite{Schechter:1980gr} where one of the ``left-handed'' neutrinos fails to pair-off
\footnote{Hybrid schemes with just one seesaw-neutrino-mass and one radiative-mass can also describe neutrino oscillations~\cite{Rojas:2018wym}.}.
The presence of a massless neutrino has a very simple and clear implication concerning neutrinoless double beta ($0\nu 2 \beta$) decay.
Indeed, if the lightest neutrino is massless, there is only one physical Majorana phase, and the effective mass parameter characterizing the amplitude for $0\nu 2 \beta$ decay has a lower limit, 
even for a normal neutrino mass ordering, as currently preferred by oscillation data~\cite{deSalas:2017kay,deSalas:2018bym}.
This is in sharp contrast to the standard three-massive-neutrino-scenario in which there can be in general a destructive interference amongst the three light neutrinos
(such cancellation in $0\nu 2 \beta$ decay may also be avoided in the three-massive-neutrino case in the presence of specific family symmetries~\cite{Dorame:2011eb,Dorame:2012zv,King:2013hj}). \\[-.3cm]

Our minimal scenario is a generalization of the scotogenic model initially proposed in~\cite{Ma:2006km}.
The basic idea of this approach is that dark matter is the mediator of neutrino mass generation, and that the same $\mathbb{Z} _{2}$ symmetry that makes the neutrino mass to have radiative origin 
also serves to stabilize dark matter.
We reexamine the Singlet + Triplet Scotogenic Model extension proposed in~\cite{Hirsch:2013ola}, which generalizes the original idea introduced in~\cite{Ma:2006km}, making its phenomenology viable and substantially richer.
  The presence of Singlet and Triplet fermions in such a scotogenic model extension automatically leads to two oscillation lengths associated to solar and atmospheric oscillations, that can be traced
  to each of the dark fermion types, leaving a massless neutrino.
  Compared to the simple scotogenic model, the unwanted spontaneous breaking of the $\mathbb{Z}_{2}$ parity symmetry~\cite{Merle:2015ica} can be naturally avoided due to the effect of the new couplings, as discussed in~\cite{Merle:2016scw}.
  Two dark matter candidates can be envisaged within the scotogenic framework, either the lightest dark fermion or the isodoublet dark scalar boson.
  Either possibility corresponds to that of a weakly interacting massive particle (WIMP), produced thermally in the early Universe, similarly to the SM particles. Either will constitute what is called
  cold dark matter.
  Following~\cite{Diaz:2016udz} here we focus on the case of scalar WIMP dark matter.  The fermionic dark matter case mimics neutralino dark matter in supersymmetry~\cite{Hirsch:2013ola} and has been
  recently re-visited in Refs.~\cite{Choubey:2017yyn,Restrepo:2019ilz,Hagedorn:2018spx}.  As we will see, scalar WIMP dark matter phenomenology in the Singlet + Triplet Scotogenic Model provides a rich scenario, sharing
  common features with other dark matter models such as the Inert Higgs Doublet Model~\cite{Deshpande:1977rw,Barbieri:2006dq,Diaz:2015pyv,Honorez:2010re,LopezHonorez:2006gr} and with discrete dark matter models~\cite{Hirsch:2010ru,Boucenna:2012qb}. 
In this paper we build upon previous analyses, expanding them by including a more detailed phenomenological study of the scalar dark matter candidate.
From the dark matter point of view our work contains many new improvements. The updated experimental constraints, in particular from dark matter direct detection experiments like XENON-1T, cut out a relevant region of the parameter space especially in the mass range around and above 100 GeV. Then we present a novel analysis of indirect probes via $\gamma$ rays. Here we compare both current as well as future $\gamma$-ray telescopes. Finally, we perform a new collider study focusing on the $\met +$ jet (mono-jet) signal, with relevant implications for future Large Hadron Collider (LHC) searches with higher luminosity.  
Moreover, we stress an interesting feature of this scenario for neutrinoless double beta decay facilities, namely the existence of a lower bound that holds even for normal neutrino mass ordering. This is very promising for the upcoming experiments. While this feature is generic to any theory in which one of the light neutrinos is massless, we stress that it comes automatically in this model, allowing for a testable connection between neutrino physics and dark matter.

  The paper is organised as follows. In section \ref{sec:model} we introduce the model, detailing the new fields and new interactions present.
  We describe in detail the scalar and fermionic sector, as well as the radiative neutrino mass generation, emphasizing that the lightest neutrino is massless and discussing the resulting lower bound
  for neutrinoless double beta decay.
  Section \ref{Sec:analysisDM} describes the numerical analysis used to study the dark matter sector of the model, listing the main constraints included. Here we have assumed that the dark matter
  is a scalar particle.
  In Section \ref{Sec:DMpheno} we present the main results concerning the relic scalar dark matter density, direct and indirect detection. In section \ref{sec:lhcDM} we deal
  with the implications for the LHC searches, taking into account the main results of the previous section.
  We focus on the $\met +$ jet (mono-jet) signal.  Finally, we give our conclusions in section \ref{sec:Conclusions}.


\section{The Singlet + Triplet Scotogenic Model}
\label{sec:model}

In this section we will review the Singlet + Triplet Scotogenic Model. This generalization of the scotogenic
model~\cite{Ma:2006km} was proposed in~\cite{Hirsch:2013ola} and
further studied in several
papers~\cite{Merle:2016scw,Diaz:2016udz,Choubey:2017yyn,Restrepo:2019ilz}.
In addition to the SM gauge symmetry there is a discrete
$\mathbb{Z} _{2}$ symmetry, whose role is to make the lightest
$\mathbb{Z} _{2}$-odd or ``dark'' particle stable and to ensure the
radiative generation of neutrino masses.  The SM particle content is
augmented by the inclusion of a Majorana fermion triplet $\Sigma$ and
a Majorana fermion singlet $F$, both odd under the $\mathbb{Z} _{2}$
symmetry.  Moreover, the model includes a new scalar doublet $\eta$
--- odd under the $\mathbb{Z} _{2}$ symmetry, which does not acquire a
vacuum expectation value (VEV) --- and a triplet scalar $\Omega$,
which allows for the mixing of the neutral parts of the new
fermions. This triplet scalar field has a zero hypercharge and it is
even under the $\mathbb{Z} _{2}$ symmetry, thus, its neutral component
can acquire a nonzero VEV. The full particle content of the model is
given in Table~\ref{tab:fields_charges}, with the corresponding
charge assignment under the different symmetry groups. 
\begin{table}[htb!]
\centering
\begin{tabular}{| c | c | c | c | c | c | c | c | }
\hline
& \multicolumn{3}{ c |  }{\quad Standard Model \quad} &  \multicolumn{2}{ c | }{\quad new fermions \quad}  & \multicolumn{2}{ c |}{\quad new scalars \quad}  \\
\hline
    &   $L$   & $e$   & $\phi$   & $\Sigma$ &    $F$    & $\eta$ & $\Omega$ \\
\hline 
Generations & \quad  3 & \quad  3 & \quad  1 & \quad  1 & \quad  1 & \quad  1 & \quad 1 \\
\hline   
\hline 
$\rm SU(3)_C$  & \quad   1    & \quad  1    & \quad   1     & \quad    1     &   \quad  1    & \quad    1    &   \quad  1     \\                                                                      
$\rm SU(2)_L$  & \quad   2    & \quad  1    &  \quad   2     & \quad    3     &   \quad  1    & \quad    2    &   \quad  3     \\
$\rm U(1)_Y$        & \quad  -1    & \quad -2    & \quad    1     & \quad    0     &   \quad  0    & \quad    1    &  \quad   0     \\
$\mathbb{Z} _{2}$      & \quad  $+$   & \quad $+$   & \quad   $+$    & \quad   $-$    & \quad    $-$   & \quad   $-$   &  \quad  $+$    \\
\hline    
$\rm L$        & \quad   1    & \quad  1    &   \quad  0     & \quad    0     & \quad    0    & \quad    -1    &  \quad   0     \\
\hline 
\end{tabular}
\caption{Particle content and quantum numbers of the Singlet + Triplet Scotogenic Model.
The charge assignments of the fields under the global Lepton Number symmetry ($\rm L$) are also shown.}
\label{tab:fields_charges}
\end{table}

Taking into account the new fields and symmetries of the model, the relevant terms of the Lagrangian read
\begin{eqnarray}
\label{eq:laginteraction}
\mathcal{L} & \subset & -Y^{\alpha \beta} L_\alpha e_\beta \phi - Y_{F}^{\alpha}(\bar{L}_{\alpha}\tilde{\eta})F- Y^{\alpha}_{\Sigma}\bar{L}_{\alpha}^{c}\Sigma^{\dag}\tilde{\eta} -Y_{\Omega} {\rm Tr} \left[\bar{\Sigma}\Omega \right]F \nonumber\\
&-& \frac{1}{2} M_{\Sigma} {\rm Tr} \left(\overline{\Sigma}^{c} \Sigma \right) - \frac{M_F}{2} \overline{F^{c}} F  + h.c.\end{eqnarray}
where $\tilde{\eta}=i\sigma_2 \eta^{\ast}$. The first Yukawa term $Y^{\alpha \beta}$ is the \sm interaction for leptons, which we can assume to be diagonal in flavor (Greek indices stand for family indices).

\subsection{Scalar sector}

The scalar potential $\mathcal{V} $ invariant under the  SU(2)$\times$U(1)$\times \mathbb{Z} _{2}$ symmetry 
is
\begin{eqnarray}
\mathcal{V} &=&-m_{\phi}^2\phi^{\dag}\phi+m_{\eta}^{2}\eta^{\dag}\eta -\frac{m_{\Omega}^{2}}{2}\rm Tr \left(\Omega^{\dag}\Omega\right) \nonumber\\
&+& \frac{\lambda_1}{2}\left(\phi^{\dag}\phi\right)^{2}+\frac{\lambda_2}{2}\left(\eta^{\dag}\eta\right)^{2}+\frac{\lambda_3}{2}\left(\phi^{\dag}\phi\right)\left(\eta^{\dag}\eta\right) + \lambda_{4}\left(\phi^{\dag}\eta\right)\left(\eta^{\dag}\phi\right)+\frac{\lambda_5}{2}\left[ \left(\phi^{\dag}\eta\right)^2 + \left(\eta^{\dag}\phi\right)^2 \right] \nonumber \\  
&+& \mu_{1}\phi^{\dag}\Omega\phi + \mu_{2}\eta^{\dag}\Omega\eta \nonumber\\
&+&\frac{\lambda_{1}^{\Omega}}{2}\left(\phi^{\dag}\phi \right) \rm Tr\left(\Omega^{\dag}\Omega\right) +\frac{\lambda_{2}^{\Omega}}{4}\left[ \rm Tr\left(\Omega^{\dag}\Omega\right) \right]^2
+\frac{\lambda_{\eta}^{\Omega}}{2}\left(\eta^{\dag}\eta \right) \rm Tr\left(\Omega^{\dag}\Omega\right),
\label{eq:scalarpot}
\end{eqnarray}
where we make the conservative assumption that $m_{\phi}^2, m_{\eta}^2$ and $m_{\Omega}^2$ are all positive, so that the spontaneous electroweak symmetry breaking will be driven by $\phi$ and (sub-dominantly) by the neutral component 
of $\Omega$, while $\eta$ cannot acquire a VEV. 
Notice that we are using the standard $2 \times 2$ matrix notation for the $\rm SU(2)_L$ triplets:  
\begin{equation}
\Sigma = \left( \begin{array}{cc}
\frac{\Sigma^0}{\sqrt{2}} & \Sigma^+ \\
\Sigma^- & -\frac{\Sigma^0}{\sqrt{2}}
\end{array} \right) \, , \quad \Omega = \left( \begin{array}{cc}
\frac{\Omega^0}{\sqrt{2}} & \Omega^+ \\
\Omega^- & -\frac{\Omega^0}{\sqrt{2}}
\end{array} \right) \, . \label{eq:triplets}
\end{equation}
The other couplings appearing in Eq.~\ref{eq:scalarpot} are constrained by a number of theoretical considerations. First, they
must comply with the condition that the potential is bounded from below in order to have a stable minimum. This requirement leads to the
following conditions~\cite{Merle:2016scw,Diaz:2016udz} 
{\small 
\begin{eqnarray}
\label{eq:Vconditions}
\lambda_{1}\geq 0,\,\,\,\,\,\,\,\lambda_{2}\geq 0,\,\,\,\,\,\,\,\lambda_{2}^{\Omega}\geq 0,~\\
\lambda_{3}+\sqrt{\lambda_{1}\lambda_{2}}\geq 0, \,\,\,\,\,\,\,\lambda_{3}+\lambda_{4}-|\lambda_{5}|+\sqrt{\lambda_{1}\lambda_{2}}\geq 0,~\\
\lambda_{1}^{\Omega}+\sqrt{2\lambda_{1}\lambda_{2}^{\Omega}}\geq0,\,\,\,\,\,\,\,\lambda_{\eta}^{\Omega}+\sqrt{2\lambda_{2}\lambda_{2}^{\Omega}}\geq0,~\\
\sqrt{2\lambda_{1}\lambda_{2}\lambda_{2}^{\Omega}}+\lambda_{3}\sqrt{2\lambda_{2}^{\Omega}}+\lambda_{1}^{\Omega}\sqrt{\lambda_{2}}+ \lambda_{\eta}^{\Omega}\sqrt{\lambda_{1}}+\sqrt{\left(\lambda_{3}+\sqrt{\lambda_{1}\lambda_{2}}\right)\left(\lambda_{1}^{\Omega}+2\sqrt{\lambda_{1}\lambda_{2}^{\Omega}}\right)\left( \lambda_{\eta}^{\Omega}+\sqrt{\lambda_{2}\lambda_{2}^{\Omega}}\right)}\geq 0.~ \,
\end{eqnarray}
}%

It is worth noticing that while these conditions ensure that $\mathcal{V} $ is consistently bounded from below at the electroweak
scale, the running of the RGEs may lead to breaking of the $\mathbb{Z}_{2}$ symmetry at some higher energy scale.  Another theory restriction
 comes from the requirement that the expansion of the potential $\mathcal{V}$ around its minimum must be perturbatively valid. 
In order to ensure this we require that the scalar quartic couplings in Eq.~\ref{eq:scalarpot} are$~\lesssim 1$. 

As mentioned before, $\eta$ does not acquire a VEV and therefore the symmetry breaking is driven only by $\phi$ and $\Omega$, which have non-zero vevs: 
\begin{equation}
\label{eq:vevs}
\langle \phi^0 \rangle = v_\phi,~~~~\langle \Omega^0 \rangle = v_\Omega
\end{equation}

The fields $\eta$, $\phi$ and $\Omega$ are written as follows
\begin{equation}
\eta=\left(\begin{array}{c}
\eta^{+}  \\
(\eta_{R}+i\eta_{I})/\sqrt{2} 
\end{array}\right),~~ \phi=\left(\begin{array}{c}
\varphi^{+}  \\
(h_{0}+v_{\phi}+i\psi)/\sqrt{2} 
\end{array}\right),~~
\Omega=\left(\begin{array}{cc}
(\Omega_{0}+v_{\Omega})/\sqrt{2} &\Omega^{+}  \\
 \Omega^{-} &-(\Omega_{0}+v_{\Omega})/\sqrt{2} 
\end{array}\right),
\end{equation}
where $\Omega_{0}$ is real and does not contribute to the CP-odd
scalar sector.  After symmetry breaking there are three charged scalar
fields (only two of which are physical, since one is absorbed by the W
boson), plus three CP-even neutral fields, and one physical CP-odd
neutral field (since the other is absorbed by the Z boson).  The VEVs
in Eq.~\ref{eq:vevs} are restricted by the following tadpole equations
or minimization conditions 
\begin{eqnarray}
\label{eq:tadpoles}
\frac{\partial\mathcal{V}}{\partial \phi}  &=&v_{\phi}\left(-m_{\phi}^{2}+\frac{1}{2}\lambda_{1}v_{\phi}^{2} - \frac{\mu_1}{2} v_{\Omega}  + \frac{\lambda_1^\Omega}{4} v_\Omega^2 \right) = 0,\\
\frac{\partial\mathcal{V}}{\partial \Omega} &=&- 2m_{\Omega}^{2}v_{\Omega} + \lambda_{2}^{\Omega}v_{\Omega}^3 + v_{\phi}^{2} \left(\lambda_{1}^{\Omega} v_{\Omega}  -\mu_1\right) = 0,
\nonumber
\end{eqnarray}
which we solve for $m_{\phi}^{2}$ and $m_{\Omega}^{2}$. 

As for the neutral sector,  the mass matrix of the CP-even (and $\mathbb{Z}_{2}-$even) neutral scalars in the basis $(\phi_{0}, \Omega_{0})$ reads
\begin{equation} \label{neutral}
\mathcal{M}^2_{h} = \left( 
\begin{array}{cc}
\Big(- m_\phi^2  + \frac{3}{2} \lambda_1 v_\phi^{2}  + v_\Omega \Big(- \mu_1 + \frac{\lambda^\Omega_1}{4} v_\Omega \Big)\Big) &\frac{1}{2} v \Big(\lambda^\Omega_1 v_\Omega  - 2\mu_1 \Big)\\ 
\frac{1}{2} v_\phi \Big(\lambda^\Omega_1 v_\Omega  - 2\mu_1 \Big) & \Big(-\frac{1}{2} m_\Omega^2  + \frac{3}{4} \lambda^\Omega_2 v_{\Omega}^{2}  + \frac{1}{4} \lambda^\Omega_1 v_\phi^{2} \Big)\end{array} 
\right) .
 \end{equation} 
 The lightest of the neutral scalar mass eigenstates is
 identified with the SM Higgs boson, $h^0$ with mass $\sim 125$ GeV,
 while the second state, $H$ is a heavier neutral scalar. 

On the other hand, the mass matrix for the charged scalars is given as
{
\begin{equation} 
\mathcal{M}^2_{H^\pm} = \left( 
\begin{array}{cc}
\frac{1}{4} \Big(2 \lambda_1 v_\phi^{2}  -4 m_\phi^2  + v_\Omega \Big(2 \mu_1  + \lambda^\Omega_1 v_\Omega \Big)\Big) & \frac{\mu_1 v_\phi}{\sqrt{2}} \\ 
 \frac{\mu_1 v_\phi}{\sqrt{2}}   &\frac{1}{2} \Big(-2 m_\Omega^2  + \lambda^\Omega_1 v_\phi^{2}  + \lambda^\Omega_2 v_\Omega^{2} \Big)\end{array} 
\right) .
 \end{equation} 
 }

 Note that, while the Z boson gets its longitudinal component only
 from the Higgs doublet $\phi$ and not from the triplet (because
 $\Omega^{0}$ is real), the charged Goldstone boson is instead a
 linear combination of $\phi^{+}$ and $\Omega^{+}$. The VEV of
 $\Omega$ will then contribute to the W boson mass, thus leading to an
 upper limit $v_\Omega \lesssim 5$ GeV~\cite{Gunion:1989ci,Gunion:1989we}: 
\begin{eqnarray}
m_Z^2 &=& \frac{1}{4} \left(g^2 + g'^2 \right) v_\phi^2 \, , \nonumber \\
m_W^2 &= & \frac{1}{4} \, g^2 \left( v_\phi^2 + 4 \, v_\Omega^2 \right) \, .
\end{eqnarray}

The mass of the new charged scalar bosons will be 
{
\begin{eqnarray}
m_{H^\pm}^{2}&=&\mu_{1}\frac{(v^{2}_{\phi}+2v^{2}_{\Omega})}{2v_{\Omega}},\\
m_{\eta^{\pm}}^{2}&=&m_{\eta}^{2}+\frac{1}{2}\lambda_{3}v_{\phi}^{2}+\frac{1}{\sqrt{2}}\mu_{2}v_{\Omega}+ \frac{1}{2} \lambda_{\eta}^{\Omega} v_{\Omega}^{2}.
\label{scalarcharged}
\end{eqnarray}
}

Because of the conservation of the $\mathbb{Z}_{2}$ symmetry, the
$\mathbb{Z}_{2}$-odd scalar field $\eta$ does not mix with any other
scalar. It proves convenient to write it in terms of its CP-even and
CP-odd components: 
$$
\eta^0 = \frac{(\eta_{R}+i\eta_{I})}{\sqrt{2}}.
$$
The physical masses of the neutral $\eta$ field are easily determined as
\begin{eqnarray}\label{eta0}
m_{\eta_{R}}^{2}&=&m_{\eta}^{2}+\frac{1}{2}(\lambda_{3}+\lambda_{4}+\lambda_{5})v_{\phi}^{2}+\frac{1}{2}\lambda_{\eta}^{\Omega} v_{\Omega}^{2} -\frac{1}{\sqrt{2}}\mu_{2}v_{\Omega},\\
m_{\eta_{I}}^{2}&=&m_{\eta}^{2}+\frac{1}{2}(\lambda_{3}+\lambda_{4}-\lambda_{5})v_{\phi}^{2}+\frac{1}{2}\lambda_{\eta}^{\Omega} v_{\Omega}^{2} -\frac{1}{\sqrt{2}}\mu_{2}v_{\Omega}.
\label{eq:metaRI}
\end{eqnarray}

The difference $m_{\eta_{R}}^{2} -m_{\eta_{I}}^{2}$ depends only on
the parameter $\lambda_{5}$ which, as we shall see in the next
paragraph, is also responsible for the smallness of neutrino
masses. In the limit $\lambda_{5} \to 0$ lepton number conservation
is restored. Hence, by construction, neutrino masses are ``natural'',
in 't Hooft's sense~\cite{tHooft:1979rat}, i.e. they are
``symmetry-protected''. Moreover, the $\mathbb{Z}_{2}$ symmetry
conservation also makes the lightest of the two eigenstates
$\eta_{R,I}$ a viable dark matter candidate, as we will discuss in
detail in section~\ref{Sec:DMpheno}.

\subsection{Fermionic sector}

Concerning the fermionic sector, the new triplet scalar $\Omega$
allows for a mixing between the singlet and triplet fermion fields
--- $F$ and $\Sigma$ --- through the Yukawa coupling $Y_\Omega$, as
shown in Eq.~\ref{eq:laginteraction}: The mass matrix for the new
fermions, in the basis $(\Sigma_{0},F)$ is given as  
\begin{equation}
\label{eq:mixmatrixfer}
\mathcal{M}_{\chi}=\left(\begin{array}{cc}
M_\Sigma  & \frac{1}{\sqrt{2}}Y_{\Omega}v_{\Omega}\\
\frac{1}{\sqrt{2}}Y_{\Omega}v_{\Omega} & M_{F}
\end{array}\right).
\end{equation}

When the neutral part of $\Omega$ acquires a VEV $v_{\Omega} \neq 0$,
the diagonalization of the mass matrix Eq.~\ref{eq:mixmatrixfer} leads
to eigenstates with the following masses (at tree level): 
\begin{eqnarray}
m_{\chi}^{\pm}&=&M_{\Sigma}, \label{mchi1}\\
m_{\chi_{1}^{0}}&=&\frac{1}{2}\left(M_{\Sigma}+M_{F}-\sqrt{\left(M_{\Sigma}-M_{F}\right)^{2}+4(2Y_{\Omega}v_{\Omega})^2}\right), \label{mchi2}\\\
m_{\chi_{2}^{0}}&=&\frac{1}{2}\left(M_{\Sigma}+M_{F}+\sqrt{\left(M_{\Sigma}-M_{F}\right)^{2}+4(2Y_{\Omega}v_{\Omega})^2}\right),\\\
\tan(2\theta)&=&\frac{4Y_{\Omega}v_{\Omega}}{M_{\Sigma}-M_{F}},
\end{eqnarray}
where $\theta$ is the mixing angle between the neutral fermion triplet
$\Sigma_{0}$ and $F$, $M_{\Sigma}$ and $M_F$ are the Majorana mass
terms for the triplet and the singlet, respectively.  Although we will
not consider this case here, it is interesting to notice that the
lightest neutral eigenstate, $\chi_{1}^{0}$ or $\chi_{2}^{0}$ may also
play the role of the dark matter~\cite{Hirsch:2013ola}, for more
recent analyses see~\cite{Choubey:2017yyn,Restrepo:2019ilz}.

\subsection{Neutrino masses} 

\begin{figure}[!htb]
\centering
\includegraphics[scale=0.32]{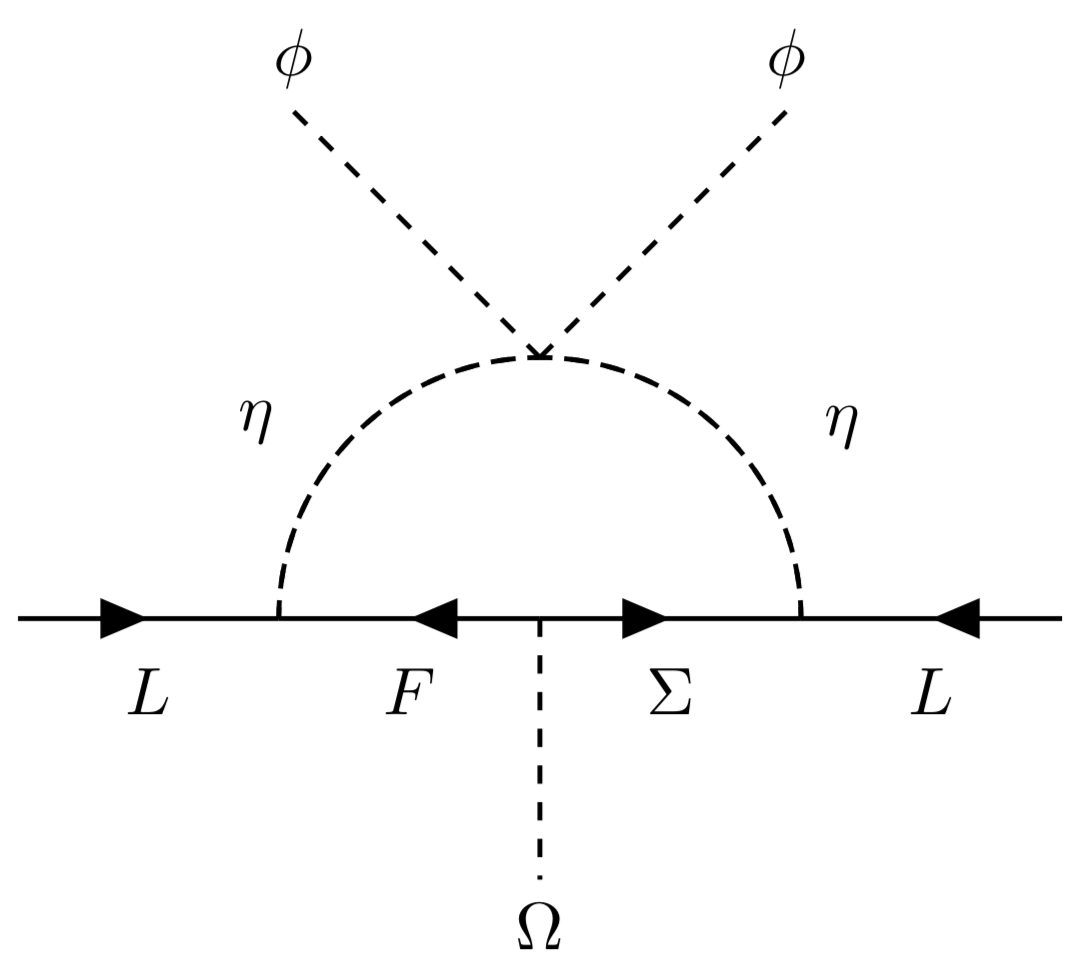}
\caption{``Scotogenic" neutrino masses. After electroweak symmetry
  breaking the SM-like Higgs acquires a VEV $\vev{\phi_0}$. }
\label{fig:nu_loop}
\end{figure}

The previous subsection has been dedicated to the spectrum of the new
fermions.  Let us now comment on neutrino masses. By construction, in
the {\textit{scotogenic}} approach, the dark matter candidate acts as
a messenger for neutrino mass generation. Since the $\mathbb{Z}_{2}$
symmetry is exact, all vertices including new particles must contain
an even number of $\mathbb{Z}_{2}$-odd fields.  For this reason
neutrinos cannot acquire a tree-level mass term, their masses arising
only at the loop level as portrayed in Fig.~\ref{fig:nu_loop}.  

The relevant interactions for the generation of neutrino masses arise
from equations~\ref{eq:laginteraction} and~\ref{eq:scalarpot}.
The expression for the neutrino mass matrix is~\cite{Hirsch:2013ola,Merle:2016scw,Diaz:2016udz}
\begin{eqnarray}
\label{eq:neutrinomass}
\mathcal{M}_{\alpha\beta}^{\nu}&=&\sum\limits_{\sigma=1,2}\frac{Y^\nu_{\alpha\sigma}Y^\nu_{\beta\sigma}}{32\pi^2} \mathcal{I}_\sigma(m_{\chi_\sigma}^2,m_{\eta_R}^2,m_{\eta_I}^2) \nonumber \\
&=&\sum\limits_{\sigma=1,2}\frac{Y^\nu_{\alpha\sigma}Y^\nu_{\beta\sigma}}{32\pi^2}m_{\chi_\sigma}\left(\frac{m_{\eta_R}^2}{m_{\eta_R}^2-m_{\chi_\sigma}^2}\ln \left(\frac{m_{\eta_R}^2}{m_{\chi_\sigma}^2}\right)-\frac{m_{\eta_I}^2}{m_{\eta_I}^2-m_{\chi_\sigma}^2}\ln \left(\frac{m_{\eta_I}^2}{m_{\chi_\sigma}^2}\right)\right),
\end{eqnarray}
where $\alpha$ and $\beta$ are generation indices
($\alpha,\beta=1,2,3$), $m_{\chi_\sigma}$ are the masses of the
$\chi^{0}_{1,2}$ fermion fields and $Y^\nu_{\alpha\beta}$ are the new
neutrino Yukawa couplings introduced as a $3\times2$
matrix~\footnote{The new fermions $\Sigma$ and $F$ match exactly the
  minimum set needed to describe neutrino oscillations. Indeed, if
  only one of them is present, the neutrino mass matrix would have
  only one nonzero eigenvalue, hence unable to account for the solar
  and atmospheric scales.},
 \begin{eqnarray}
 \label{eq:nuyuk}
Y^\nu&=& \begin{pmatrix}
Y_{\Sigma}^1 & Y_{F}^1\\
Y_{\Sigma}^2 & Y_{F}^2\\
Y_{\Sigma}^3 & Y_{F}^3 
\end{pmatrix}
\cdot
V(\theta) \,.
\end{eqnarray}   

The matrix $V(\theta)$ is a $2 \times 2$ orthogonal matrix that
diagonalizes the fermionic mass matrix $\mathcal{M}_\chi$ given in
Eq.~(\ref{eq:mixmatrixfer}). As already noticed before, in the limit
$\lambda_5 \to 0$ the two eigenstates $m_{\eta_R}$ and $m_{\eta_I}$
are degenerate, hence neutrino masses are zero and the lepton number
symmetry is restored.
This limit would correspond to an exact cancellation
between the $\eta_R$ and $\eta_I$ loops.  The expression
$\mathcal{I}_\sigma(m_{\chi_\sigma}^2,m_{\eta_R}^2,m_{\eta_I}^2)$ in
Eq.~\ref{eq:neutrinomass} involves differences of Passarino-Veltman
functions $B_0$~\cite{Passarino:1978jh}, evaluated in the limit of
vanishing external momentum.  

We can then rewrite Eq.~\ref{eq:neutrinomass} more compactly as:
\begin{equation}
\mathcal{M}_{\alpha\beta}^{\nu} = Y^\nu_{\alpha\beta} v_{\phi}\cdot\frac{\mathcal{F}}{v_{\phi}^{2}}\cdot Y^{\nu, T}_{\alpha\beta} v_{\phi}\sim m_{D}\frac{1}{M_{R}}m_{D}^{T},
\end{equation}
where 
\begin{equation}
\mathcal{F} = \left( \begin{array}{cc}
\frac{\mathcal{I}_1}{32 \pi^2 } & 0 \\
0 & \frac{\mathcal{I}_2}{32 \pi^2 } 
\end{array} \right) \,  .
\label{eq:Fnumass}
\end{equation}
This recalls the structure of the standard type-I seesaw neutrino mass
relation, with the Dirac mass term given by
$Y^\nu_{\alpha\beta} v_{\phi}$ and
$M_{R}^{-1} = \frac{\mathcal{F}}{v_{\phi}^{2}}$ where $\mathcal{F}$
includes the loop functions.  In order to compare with the current
determination of neutrino oscillation
parameters~\cite{deSalas:2017kay}, we will apply a Casas-Ibarra
parametrization~\cite{Casas:2001sr}:
\begin{equation}
\label{eq:yvpar}
Y^\nu_{\alpha\beta} =U_{\nu}\sqrt{m_{\nu}}\rho\sqrt{\mathcal{F}}^{-1}, 
\end{equation}
where $U_{\nu}$ is the lepton mixing matrix, $m_{\nu}$ are the
neutrino masses (whose squared differences are constrained as
in~\cite{deSalas:2017kay}) and the matrix $\rho$ is an arbitrary
$2\times 3$ rotation matrix that can be parametrized
as~\cite{Hirsch:2013ola} 
\begin{eqnarray}
\label{eq: rhomatrixform}
\rho&=& \begin{pmatrix}
0 & \cos(\beta) & \pm \sin(\beta)\\
0 & -\sin(\beta) & \pm \cos(\beta)
\end{pmatrix}.
\end{eqnarray}

An interesting prediction of this model is that the lightest neutrino
is massless. This feature is reminiscent of the ``missing partner''
nature of this ``radiative'' seesaw mechanism, in which one of the
``right-handed'' fermions is missing (there is only one $\Sigma$ and
one $F$). As a consequence one of the ``left'' neutrinos can not
pair-off and hence remains massless~\cite{Schechter:1980gr}. 

\subsection{Neutrinoless double beta decay} 

Within the symmetrical parametrisation of the lepton mixing
matrix~\cite{Schechter:1980gr} the $0\nu 2 \beta$ effective mass
parameter can be neatly expressed as~\cite{Rodejohann:2011vc} 
\begin{equation}
\vev{m_{ee}}=\left|\sum_jU_{\nu, ej}^2m_j\right|=\left| \cos \theta_{12}^2 \cos \theta_{13}^2 m_1+ \sin \theta_{12}^2 \cos \theta_{13}^2 m_2e^{2i\phi_{12}}+ \sin \theta_{13}^2 m_3e^{2i\phi_{13}}\right|\,,
\end{equation}
where $m_i$ are the three neutrino masses and $\theta_{1x}$ are the
neutrino mixing angles measured in oscillation experiments.  
Note that in our case the lightest neutrino is massless ($m_1 = 0$),
so that there is only one physical Majorana phase
($\phi \equiv \phi_{12} - \phi_{13}$). Since there is currently no
restriction on its value, this phase is a free parameter.
Except for this, all other parameters are well measured in oscillation
experiments.

\begin{figure}[!hbt]
\centering
 \includegraphics[scale=0.58]{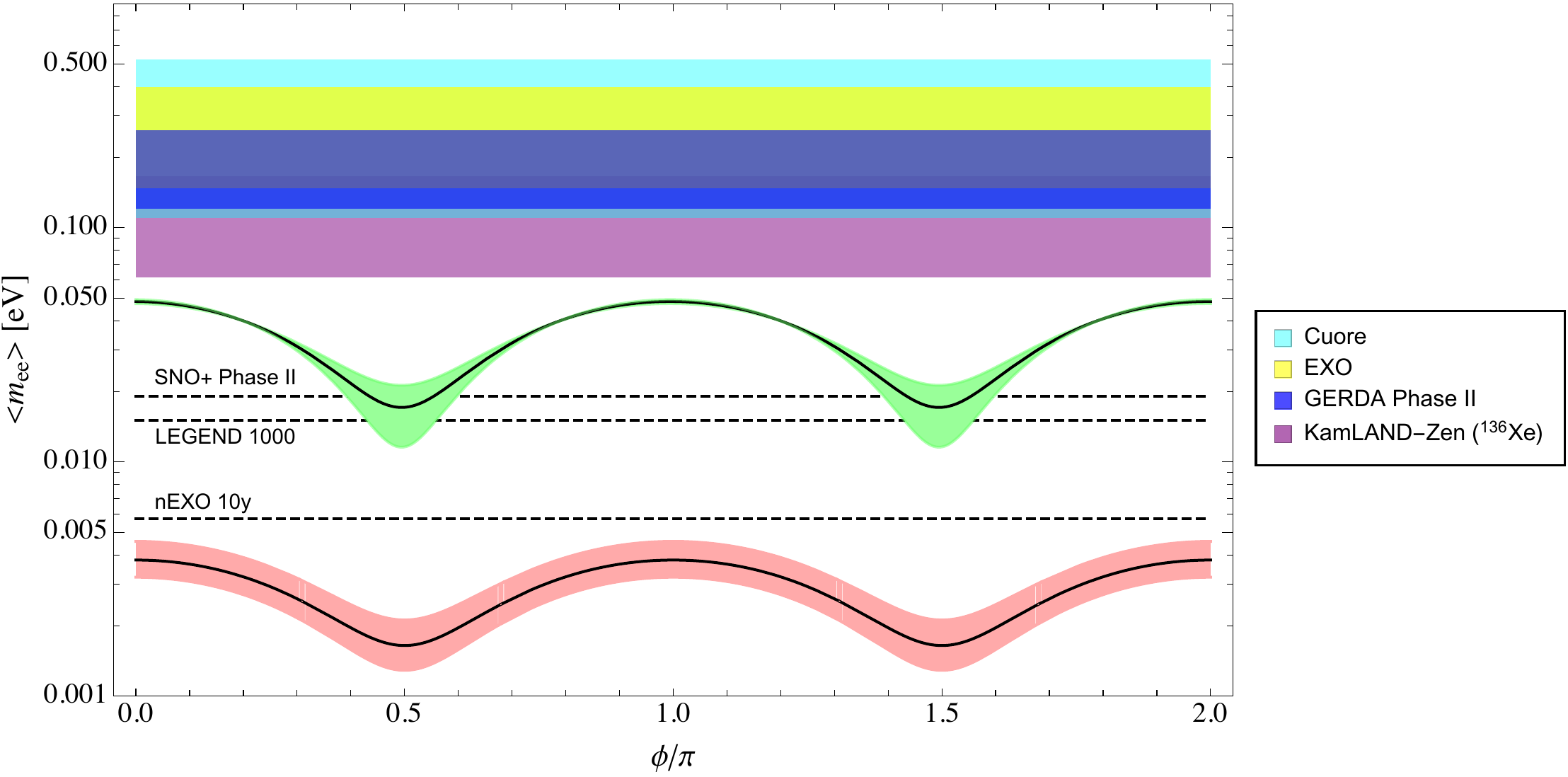}
\caption{Effective $0\nu 2 \beta$ Majorana mass parameter versus the Majorana phase. 
The pink (light green) band represents the prediction for the $3\sigma$ C.L. region allowed by current oscillation experiments for normal (inverted) mass ordering. }
\label{fig:DBD}
\end{figure}

We show in Fig.~\ref{fig:DBD} the dependence of $\vev{m_{ee}}$ on this
phase. 
One sees that, in contrast to the general case where the three active
Majorana neutrinos are massive, here the effective mass parameter
describing the $0\nu 2 \beta$ decay amplitude has a lower
limit~\cite{Reig:2018ztc,Leite:2019grf}. 

The pink (light green) band refers to the $3\sigma$ C.L. region allowed by
current oscillation experiments~\cite{deSalas:2017kay} for normal
(inverted) mass ordering. The black lines correspond to the best fit
values for both cases. 
We also show for comparison the 90\% C.L. upper limits (shaded regions) from different experiments: CUORE ($\vev{m_{\beta\beta}}<110-520$ [meV]) \cite{Alduino:2017ehq}, EXO-200 Phase II ($147-398$ [meV]) \cite{Albert:2017owj}, GERDA Phase II ($120-260$ [meV])
\cite{Agostini:2018tnm} and KamLAND-Zen ($61-165$ [meV]) \cite{KamLAND-Zen:2016pfg}
experiments. The width of these bands is mainly a reflection of the uncertainty in the relevant
nuclear matrix elements. The black dashed lines represent the most optimistic future sensitivities
for SNO+ Phase II ($19-46$ [meV])~\cite{Andringa:2015tza}, LEGEND ($15-50$ [meV])~\cite{Abgrall:2017syy} and nEXO after 10 years of data
taking ($5.7-17.7$ [meV])~\cite{Albert:2017hjq}. 

\section{Numerical analysis}  
\label{Sec:analysisDM}

We now confront the model with current (and future) observations associated both with
the primordial cosmological abundance of dark matter, as well as various phenomenological
constraints, including the experimental prospects for direct and indirect dark matter detection.

\subsection{Parameter scan}
\label{Ssec:scan}

We have developed a numerical code using Python, to perform a scan
varying randomly the main free parameters which characterize the
model.
This code is connected to some public computer tools used in particle
physics in order to examine the constraints on the model parameters
and also quantify the expected sensitivities of future experiments.
In particular, our Singlet + Triplet Scotogenic Model is first
implemented in {\tt SARAH 4.9.1}~\cite{Staub:2013tta,Staub:2015kfa},
which calculates all vertices, mass matrices, tadpole equations,
one-loop corrections for tadpoles and self-energies.
The physical particle spectrum and low-energy observables are computed
with {\tt SPheno 4.0.3}~\cite{Porod:2003um,Porod:2011nf} and {\tt
  FlavorKit}~\cite{Porod:2014xia}.
In order to perform the dark matter analysis, we use {\tt Micromegas
  5.0.2}~\cite{Belanger:2014vza} to compute the thermal component to
the dark matter relic abundance as well as the dark matter-nucleon scattering cross
sections.
For the calculation of the cross sections relevant for the collider
analysis, we have used {\tt MadGraph5}~\cite{Alwall:2014hca},
importing the UFO files generated with {\tt SARAH 4.9.1}.
Our numerical scan was performed with 60000 points, varying the input
parameters as given in table~\ref{tab:rangeS}, assuming logarithmic steps. In particular, in the
ranges of variation for the values of $m_{\eta}^{2}$ and $|\lambda_5|$,
the lower limits considered were $100$  $\rm GeV^2$ and $10^{-5}$
respectively, to ensure good behaviour for the $\mathbb{Z} _{2}$
symmetry~\cite{Merle:2016scw}.
\begin{table}[!htb]
\centering
\begin{tabular}{|c|c|}
\hline
{\bf Parameter} & {\bf Range}  \\ 
\hline  
\hline                             
$M_N $& [$5\cdot10^3,10^4$]  \rm (GeV)    \\
$M_{\Sigma} $& [$5\cdot10^3,10^4$]     \rm (GeV)     \\
$m_{\eta}^2$& [$100 , 5000$]   $\rm (GeV^2)$     \\
$\mu_{1,2} $& [$10^{-8} , 5\cdot10^3$]   \rm (GeV)     \\
$v_\Omega $& [$10^{-5} , 5$]  \rm (GeV)    \\
$|\lambda_{i}|$, $i=1...4$ & [$10^{-8} ,1$] \\
$|\lambda_{5}|$ &  $ [10^{-5} ,1$] \\
$|\lambda_{1,2}^{\Omega}|$ & [$10^{-8} ,1$] \\
$|\lambda_{\eta}^{\Omega}|$ & [$10^{-8} ,1$] \\
$|Y_\Omega|$ & [$10^{-8},1$] \\
\hline
\end{tabular}
\caption{Ranges of variation of the input parameters used in the numerical scan.}
\label{tab:rangeS}
\end{table}

This model has in principle three potentially viable dark matter
candidates: $\eta_R$, $\eta_I$ or $\chi_0$.
In the following we will fix $\lambda_5 < 0$ so as to ensure
$\eta_{R}$ to be the dark matter candidate. This choice is made for
definiteness, having in mind that the opposite case with
$\lambda_5 > 0$ and $\eta_{I}$ as the lightest neutral scalar would
also be potentially viable.
Notice that the parameters that are not shown in the table are
calculated from the ones displayed. For example, $m_{\phi}^{2}$ and
$m_{\Omega}^{2}$ are obtained from the tadpole
equations~\ref{eq:tadpoles} and $Y^\nu_{\alpha\beta} $ is calculated
via Eq.~\ref{eq:yvpar}. Note that the smallness of neutrino masses
does not preclude these Yukawas from being sizeable, since the
neutrino masses are controlled by $\lambda_5$ and they are further
suppressed by their radiative origin. 

\subsection{Constraints}
\label{Ssec:Constraints}

The presence of new particles, absent in the Standard Model, will
induce departures from the SM predictions for a number of
observables. Throughout our analysis, we take into account the
following constraints. 

\paragraph{Theoretical constraints}

As already discussed in Sec.~\ref{sec:model}, the coupling and mass
parameters appearing in the Lagrangian Eq.~\ref{eq:laginteraction} are
subject to several theoretical constraints. First of all, we must
ensure that the scalar potential is bounded from below, which we do by
applying the conditions summarised in
Eqs.~\ref{eq:Vconditions}. 
Moreover we must ensure the perturbativity of the couplings, i.e. the
scalar quartic couplings are assumed to be $\lesssim \mathcal{O} (1)$.
Another theoretical consideration concerns the validity of the
$\mathbb{Z}_{2}$ parity symmetry which is an essential ingredient of
this model. Its role is indeed twofold: it stabilises the dark matter candidate
$\eta_R$ and it justifies the one-loop radiative seesaw mechanism
which gives mass to neutrinos. 
Compared to the simple scotogenic model initially proposed
in~\cite{Ma:2006km}, the spontaneous breaking of the $\mathbb{Z}_{2}$
parity symmetry can be naturally avoided in this extension thanks to
the effect on the running of the couplings in the scalar sector
induced by the inclusion of $\mathbb{Z}_{2}$-even scalar
triplets. 
Nevertheless, even if the scalar potential is
$\mathbb{Z}_{2}$-preserving at the electroweak scale, the RGEs running
could lead to situations where the $\mathbb{Z}_{2}$ is broken at some
higher energy scale~\cite{Merle:2016scw}\footnote{In addition, finite-temperature corrections to the effective scalar potential may affect the stability of dark matter~\cite{PhysRevD.9.3320,PhysRevD.45.2933}.}. 
While a dedicated analysis of the RGE running is beyond the scope of the present work, we have not ignored this restriction. Following the
prescriptions in~\cite{Merle:2016scw} we have avoided this problem by fixing the ranges of variation of the relevant parameters ($m_\eta^2$, $\mu_2$ and $v_\Omega$) so that the
$\mathbb{Z}_{2}$ parity symmetry holds up to higher energies~\footnote{Notice that requiring the validity of the $\mathbb{Z} _{2}$ parity symmetry might be over-restrictive. Indeed, UV-completions might contain new degrees of freedom -- irrelevant for our phenomenological study -- at high energies that could prevent the dangerous breaking.}. 
Moreover, we have checked numerically that, for benchmarks we chose for the collider study in Sec.~\ref{sec:lhcDM} $m_{\eta}^2$ remains positive at all energy scales. 
Finally, although experimental constraints place no upper limit on the
mass of the heavy neutral scalar $H$, we require that its decay width
should comply with the perturbative unitarity condition, i.e.
$\frac{\Gamma_{H}}{m_{H}}\, < \, \frac{1}{2}$.

\paragraph{Neutrino oscillation parameters}

One of the main motivation of our scotogenic model is to provide an explanation to the generation of neutrino masses. To ensure this, throughout our analysis we require compatibility with the best-fit ranges of the neutrino oscillation parameters.
This is enforced via Eq.~\ref{eq:yvpar}, where the mixing angles and squared mass differences are fixed according to Ref.~\cite{deSalas:2017kay}. For simplicity, the yet unknown Dirac and
Majorana phases in $U_\nu$ are set to zero. We further assume the currently preferred normal ordering of the light neutrino masses. Interestingly, as already mentioned, this model predicts the
lightest active neutrino to be massless. 

\paragraph{Lepton flavour violation}

This model could be in principle probed through the observation of charged lepton flavour violation, for example, at high intensity muon facilities~\cite{Rocha-Moran:2016enp}.
However, the negative results of charged \lfv searches can be used to set constraints on the parameters of the model, in particular on $\lambda_5$ which controls the magnitude of the Yukawa couplings. 
We apply the most stringent limits to date on the branching fraction of some of such rare processes, namely BR$(\mu \to e \gamma) < 4.2 \times 10^{-13}$~\cite{TheMEG:2016wtm},  BR$(\mu \to e e e)  < 1. \times 10^{-12}$~\cite{Bellgardt:1987du}, CR$(\mu^-, {\rm Au} \to e^-, \rm Au) < 7 \times 10^{-13}$~\cite{Bertl:2006up}.

\paragraph{Electroweak precision tests}

The presence of new physics will affect the gauge boson self-energies, parameterised by the oblique parameters~\cite{Peskin:1991sw}. The most important constraint is expected from the $T$ parameter,
which is sensitive to the mass splitting between the neutral and charged components of the scalar fields. From the one-loop contribution to the T parameter it follows that
    $m_{\eta^\pm} - m_{\eta_R} \lesssim 140$ GeV~\cite{Abada:2018zra}.
    Given the perturbativity constraint on the relevant $\lambda_i$ parameters, this condition is automatically fulfilled in all phenomenologically viable solutions of our numerical scan.
We require consistency with electroweak precision data by requiring $v_\Omega \lesssim 5$ GeV, in order to get an adequately small deviation of the $\rho$ parameter from one~\cite{Merle:2016scw} {(and consequently a negligible tree level contribution from the triplet to the T parameter)}, namely we impose $-0.00018 \lesssim \delta \rho \lesssim 0.00096$ ($3\sigma$). Moreover, we fix the Higgs VEV $v_\phi$ in order to get the correct mass of the $W$ boson, inside its experimental range.

\paragraph{Invisible decay widths of the Higgs boson}

If the new neutral scalar masses $m_{\eta_{R, I}}$ are small enough, there can appear new invisible decay channels -- at tree level -- of the Higgs boson into the lighter stable particles.  
In the region of parameters where these new invisible decays are possible we enforce that BR$(h^0 \to \rm inv) \lesssim 24\%$~\cite{Tanabashi:2018oca}. 
At the loop level, the decay of the Higgs boson into two photons may also be modified by its coupling to the charged scalars. We require consistency at the $3\sigma$ level, that is $0.62\lesssim$ BR$(h^0 \to \gamma \gamma)/$BR$(h^0 \to \gamma \gamma)_{\rm SM} \lesssim 1.7$. 

\paragraph{Dipole moments of leptons}

At the one-loop level the charged scalars present in our model may also induce sizeable contributions to the magnetic dipole moments of leptons.  
We have required that the contributions to the anomalous muon magnetic dipole moment induced by the new physics do not exceed the
allowed discrepancy between the measured value and the one predicted within the SM~\cite{Tanabashi:2018oca}, $\Delta (a_{\mu})\,=\,a^\text{exp}_{\mu}-a^\text{SM}_{\mu}\,=\,  268(63)(43)\times 10^{-11}\,$. 
Contributions to the electric dipole moments arise instead only at the two-loop level, so they are suppressed~\cite{Abada:2018zra}. 

\paragraph{Dark matter and cosmological observations} 

In the following, we assume a standard cosmological scenario, where the dark matter candidate, the scalar $\eta_R$, was in thermal equilibrium with the SM particles in the early Universe.
If $\eta_R$ is the only candidate contributing to the cosmological dark matter, its relic density must comply with the cosmological limits for cold dark matter derived by the Planck satellite data~\cite{Ade:2015xua,Aghanim:2018eyx}: $ 0.1126 \leq \Omega_{\eta_R} h^2 \leq  0.1246$ (3$\sigma$ range).
Values of $\Omega_{\eta_R} h^2 \leq  0.1126$ are also allowed, if $\eta_R$ is a subdominant component of the cosmological dark matter and allowing for the existence of another candidate.
Moreover, our scenario can be tested at direct detection (DD) experiments, which are meant to probe the nuclear recoil in the scattering of galactic $\eta_R$ off-nuclei inside the detector. We apply the current most stringent limit on WIMP-nucleon spin-independent (SI) elastic scattering cross section, which has been set by the XENON1T experiment at LNGS \cite{Aprile:2018dbl}. 

\paragraph{Colliders} 

Existing searches for new charged particles at colliders such as LEP and LHC, already set lower limits on their masses in the region below $100$ GeV or so~\cite{Tanabashi:2018oca}.
In our analysis we apply the following limits: $m_{H^\pm} \geq 80 \,{\rm GeV}$ and $122~{\rm GeV} \leq m_{h^0} \leq  128$ GeV, the latter
to take into account numerical uncertainties.

\black
\section{Phenomenology of Scalar Scotogenic Dark Matter} 
\label{Sec:DMpheno}

In this section we collect the results of our analysis of dark matter in the Triplet + Singlet Scotogenic Model.
As already commented before, this model can harbor either fermionic or bosonic WIMP dark matter.
Detailed studies of the phenomenology of the fermionic dark matter candidate $\chi^0$~\cite{Hirsch:2013ola} have been presented in Ref.~\cite{Choubey:2017yyn,Restrepo:2019ilz}.
Here we will assume the $\mathbb{Z}_{2}$-odd scalar $\eta_R$  to be the dark matter candidate and investigate its phenomenology.
The latter has common features with those of the simplest scotogenic constructions~\cite{Ma:2006km} as well as the Inert Higgs Doublet Model~\cite{Hambye:2009pw,Honorez:2010re,Diaz:2015pyv}.

\subsection{Relic density}

\begin{figure}[!htb]
\begin{center}
 \includegraphics[scale=0.7]{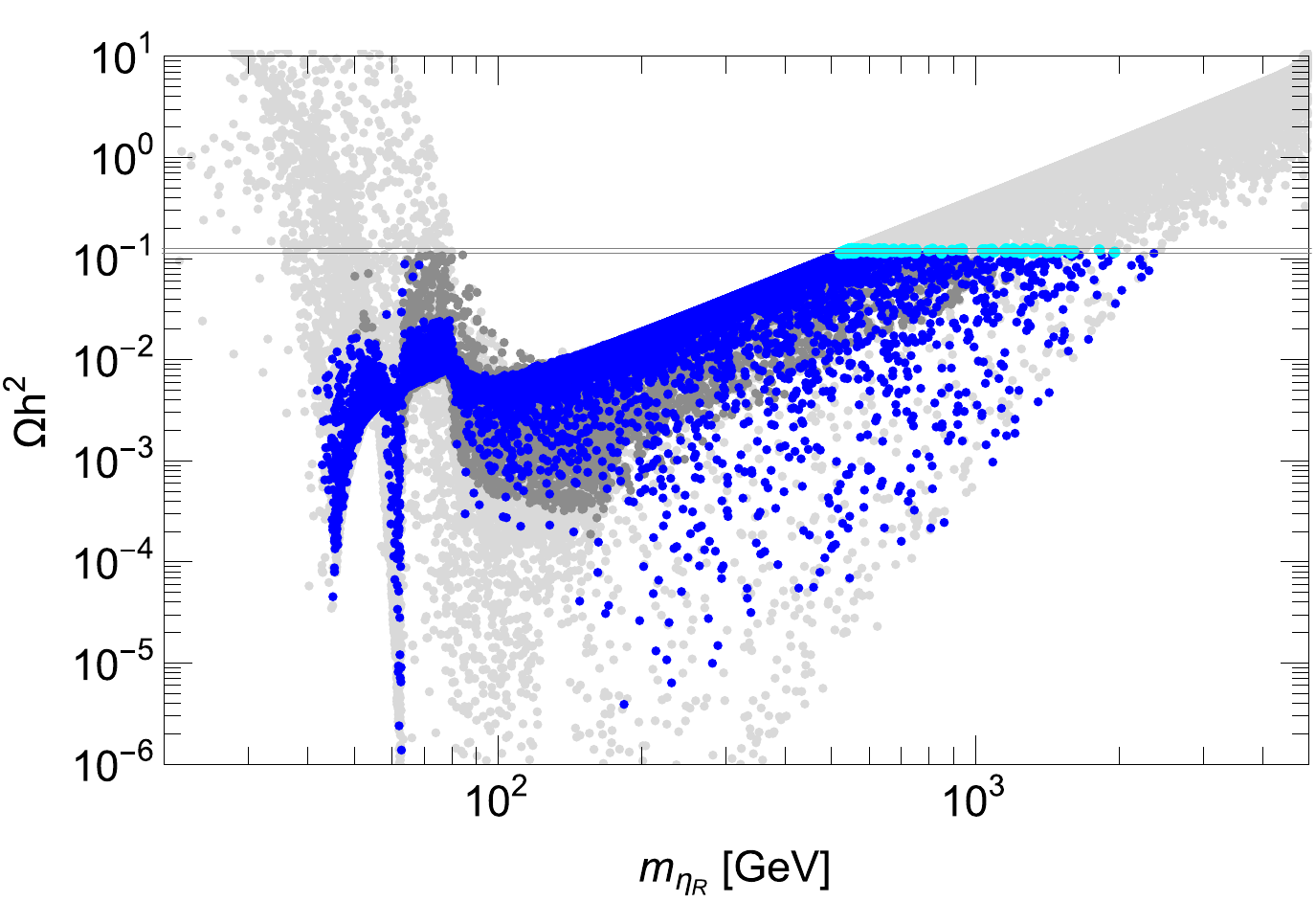}
\end{center}
\caption{Relic abundance  $\Omega_{\eta_R} h^2$ as a function of the $\eta_R$ mass. Blue points denote solutions with viable relic density, although leading to underabundant dark matter. Cyan points fall within the $3\sigma$ C.L. cold dark matter measurement by the Planck collaboration~\cite{Ade:2015xua,Aghanim:2018eyx}. Grey dots are excluded by at least one of the bounds in Sec.~\ref{Ssec:Constraints}. Dark grey points are in conflict with the current limit on WIMP-nucleon SI elastic scattering cross section set by XENON1T~\cite{Aprile:2018dbl}.}%
\label{fig:DMrelic}%
\end{figure}

We show in Fig.~\ref{fig:DMrelic} the expected dark matter relic abundance as a function of the mass of the scalar dark matter candidate $\eta_R$. The narrow black band depicts the 3$\sigma$ range for cold dark matter derived by the Planck satellite data~\cite{Ade:2015xua,Aghanim:2018eyx}. Only for solutions falling exactly in this band (cyan points) the totality of dark matter can be explained by $\eta_R$. Blue points refer to solutions where $\eta_R$ would be subdominant, and another dark matter candidate would be required. Grey points are instead excluded by any of the constraints discussed in Sec.~\ref{Ssec:Constraints}, mainly by the Planck constraint itself. Dark grey points are solutions in conflict with the current limit on WIMP-nucleon SI elastic scattering cross section set by XENON1T~\cite{Aprile:2018dbl}. The features appearing in the plot can be explained by looking in detail into the $\eta_R$ annihilation channels.
The first dip on the left depicts the $Z$-pole, that is where $m_{\eta_R} \sim M_Z/2$ and the coannihilation via s-channel $Z$ exchange becomes relevant. Similarly, the second depletion of the relic density  around $m_{\eta_R} \sim 60$ GeV corresponds to efficient annihilations via s-channel Higgs exchange.
Notice that it is likely for solutions in this dip to be in conflict with current collider limits on BR$(h^0 \to \rm inv)$.
The latter depletion is more efficient than the Z-mediated one, which is momentum suppressed. For heavier $\eta_R$ masses, quartic interactions with gauge bosons become effective and, when kinematically allowed, also two-top final states.  
Annihilations of $\eta_R$ into $W^+ W^-$ via quartic couplings are particular important at $m_{\eta_R} \gsim 80$ GeV thus explaining the third drop in the relic abundance. 
Finally, in the range $m_{\eta_R} \gsim 120$ GeV $\eta_R$ can annihilate also into two Higgs bosons. 
At even heavier $m_{\eta_R}$ the annihilation cross section drops as $\sim \frac{1}{m_{\eta_R}^2}$ and the relic density increases proportionally.
Eventually, heavy $\eta_R$ mainly annihilate into $W^+ W^-$, $h^0 h^0$, $H H$. 
We collect all the Feynman diagrams contributing to $\eta_R$ annihilations and co-annihilations in Appendix~\ref{app:relic_diagrams}.  
We may also notice that the relic abundance constraint does not put any bound on the absolute value of the $|\lambda_5|$ parameter. 
On the other hand, coannihilations with $\eta_I$ and $\eta^\pm$ may occur in all regions of the parameter space with the effect of lowering the relic abundance. 

\begin{figure}[!hbt]
\begin{center}
 \includegraphics[scale=0.7]{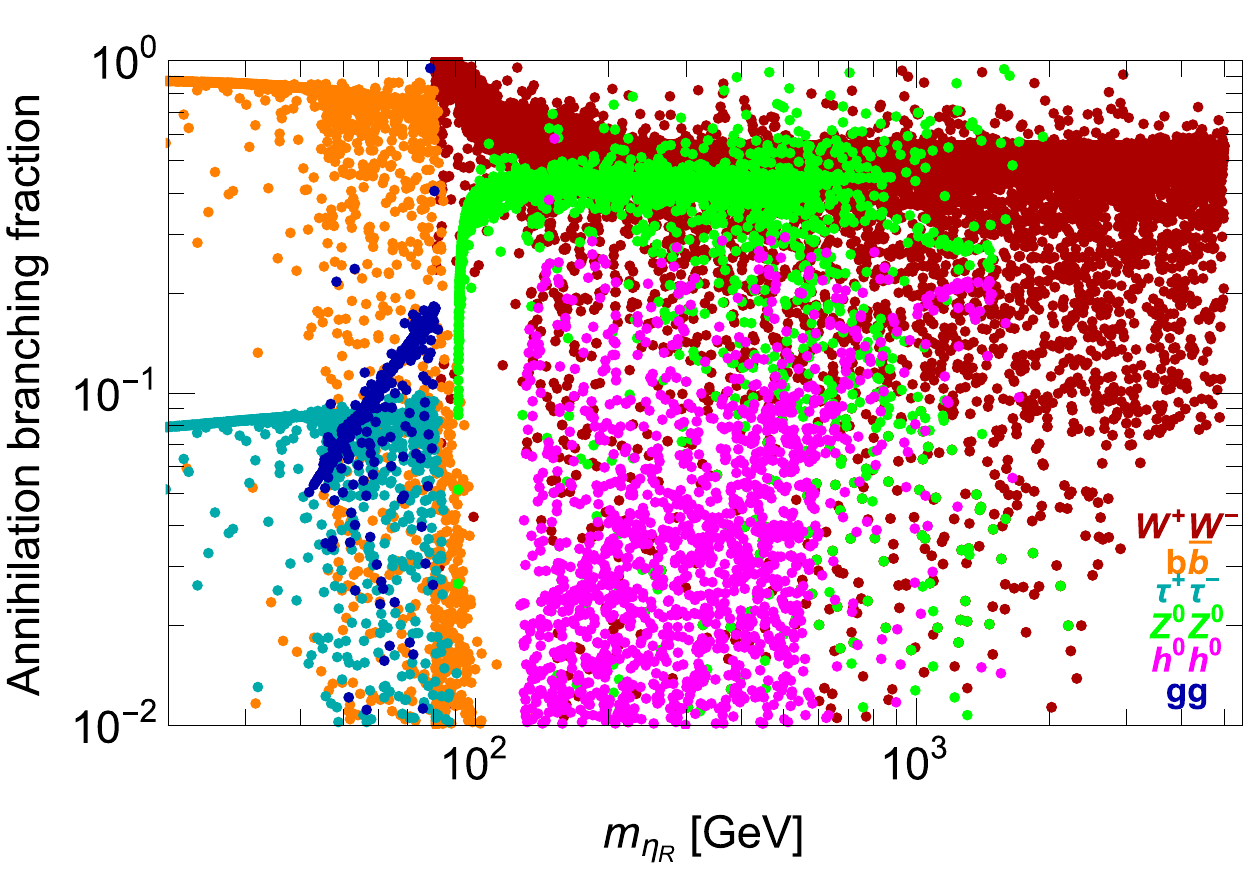}
\end{center}
\caption{Main branching fractions of the annihilation cross section of $\eta_R$ into SM final states versus the mass of $\eta_R$. Orange points refer to annihilation into $b \bar{b}$, dark cyan to $\tau^+ \tau^-$, blue to gluons, dark red to $W^+ W^-$, green to $Z^0 Z^0$ and magenta to $h^0 h^0$.}%
    \label{fig:DMBR}%
\end{figure}

We show in Fig.~\ref{fig:DMBR} the most relevant branching ratios (at tree level) for the annihilation cross section of $\eta_R$ into SM final states versus the mass of $\eta_R$, from our numerical scan.
Different kinematical regimes are visible from this figure: below $M_W$, $\eta_R$ annihilates predominantly into $b \bar{b}$, gluons or $\tau^+ \tau^-$; when the quartic coupling with W becomes kinematically accessible, $\eta_R$ annihilates mainly into $W^+ W^-$. Similarly, annihilations into $h^0 h^0$, $H H$ and $Z^0 Z^0$ become relevant as soon as kinematically open. 

\subsection{Direct detection}
\label{sec:DD}

\begin{figure}[!htb]]
  \begin{center}
 \includegraphics[scale=0.6]{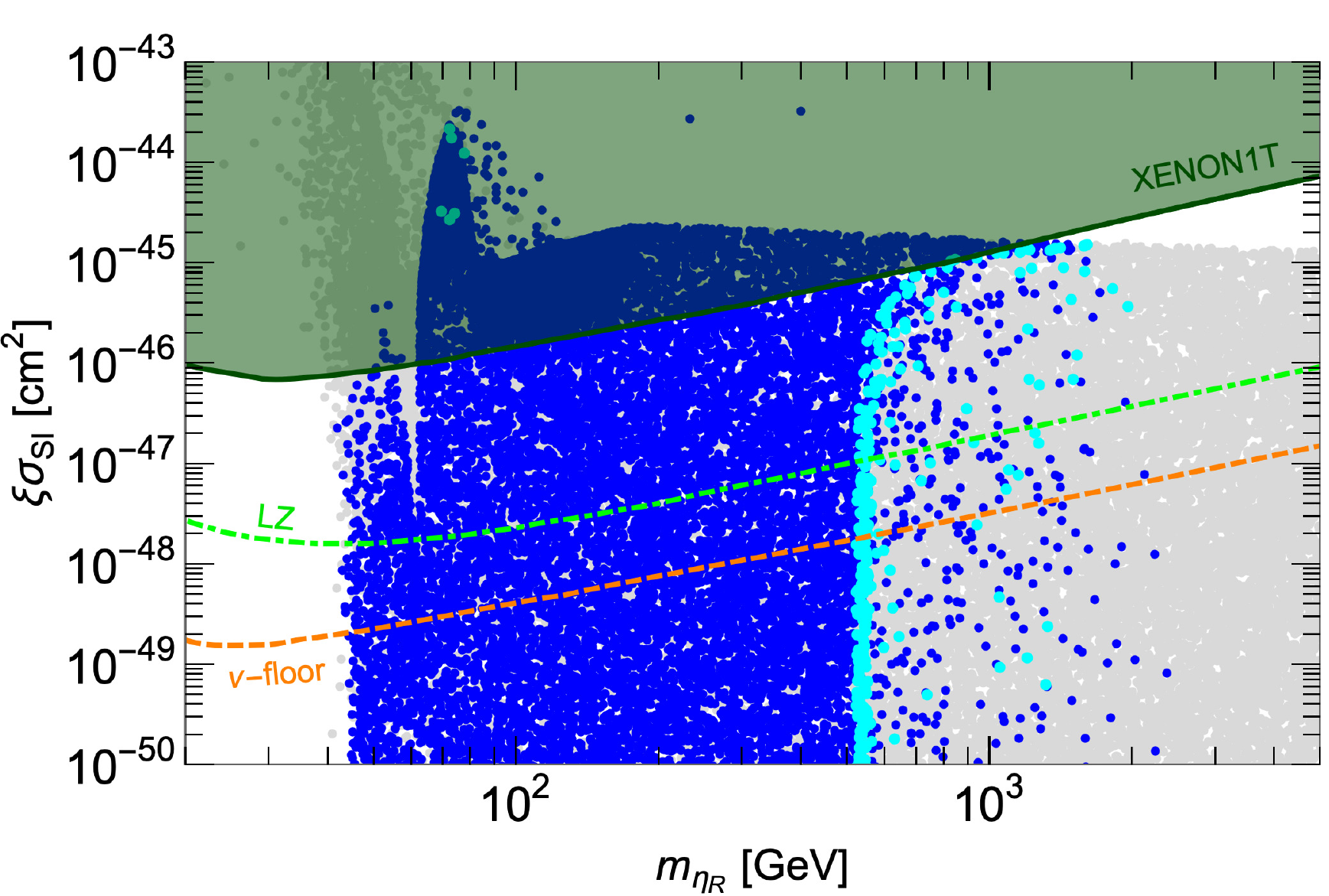}
\end{center}
\caption{Spin-independent $\eta_R$-nucleon elastic scattering cross section versus the $\eta_R$ mass. Colour code as in Fig~\ref{fig:DMrelic}. The dark green line denotes the most recent upper bound from XENON1T~\cite{Aprile:2018dbl}. The dashed orange line depicts the lower limit corresponding to the ``neutrino floor" from coherent elastic neutrino-nucleus scattering (CE$\nu$NS)~\cite{Billard:2013qya}, while the green dot-dashed one stands for the projected sensitivity for LZ~\cite{Akerib:2018lyp}.}%
    \label{fig:DMDD}%
\end{figure}

Let us discuss now the $\eta_R$ direct detection prospects.  The tree-level spin-independent $\eta_R$-nucleon interaction cross section is mediated through the Higgs and the Z portals.
The relevant Feynman diagrams for this process are summarised in Appendix~\ref{app:DD_diagrams}. 
Since the $\eta$ doublet has nonzero hypercharge, the $\eta_R$ - nucleon interaction through the Z boson would in general exceed the current constraints from direct detection experiments. 
Nevertheless, in most of the solutions, $\lambda_5$ induces a mass splitting between the CP-odd partner $\eta_I$ and $\eta_R$ such that the interaction through the Z boson is kinematically forbidden, or leads to inelastic scattering. 
The $\eta_R$-nucleon interaction via the Higgs is therefore dominant in most of the parameter space. 
As a consequence, the coupling between $\eta_R$ and the Higgs boson (which depends on the sum $\lambda_3 + \lambda_4 + \lambda_5$ and on $v_\Omega$, $\mu_2$ and $\lambda_{\eta}^{\Omega}$) turns out to be the relevant quantity controlling both this cross section and the signals at LHC that we will discuss in section~\ref{sec:lhcDM}.
We show in Fig.~\ref{fig:DMDD} the spin-independent $\eta_R$-nucleon elastic scattering cross section weighted by $\xi =  \frac{\Omega_{\eta_R}}{\Omega_{\rm Planck}}$ versus the $\eta_R$ mass.  
The color code of displayed points is the same as in Fig.~\ref{fig:DMrelic}.
The dark green plain line denotes the most recent upper bound from XENON1T~\cite{Aprile:2018dbl}.
Although we only show the most stringent up-to-date limit from XENON1T, we note that other leading liquid xenon experiments such as LUX~\cite{Akerib:2016vxi} and PandaX-II~\cite{Cui:2017nnn} can also probe the spin-independent dark matter-nucleon elastic scattering cross section for dark matter heavier than $\sim 50$ GeV.  
On the other hand, DarkSide-50~\cite{Agnes:2018ves} and DEAP-3600~\cite{Ajaj:2019imk} are less competitive for medium and high-mass WIMPs, because of their higher thresholds and lower exposures. 
Finally, we also depict as for comparison the lower limit (dashed orange line) corresponding to the ``neutrino floor'' from coherent elastic neutrino-nucleus scattering (CE$\nu$NS)~\cite{Billard:2013qya} and the projected sensitivity for LUX-ZEPLIN (LZ, green dot-dashed)~\cite{Akerib:2018lyp}. 
The extended particle content characteristic of the Singlet + Triplet Scotogenic Model in principle allows for a viable scalar dark matter candidate in a wider region of masses, compared to the simplest scotogenic or Inert Higgs Doublet models. Nevertheless, because of current experimental constraints, most of the new allowed solutions with a relic abundance within the $3\sigma$ C.L. cold dark matter measurement by the Planck collaboration~\cite{Ade:2015xua,Aghanim:2018eyx} lie in a tight vertical region around  $m_{\eta_R} \sim 500-600$ GeV. Lighter $\eta_R$ lead to viable dark matter, although under-abundant, hence it would then require the existence of an additional dark matter candidate. 

Finally, it is worth commenting on how the phenomenology of $\eta_R$ dark matter compares to that of the scalar dark matter in the simple Scotogenic Model \cite{Ma:2006km}.
While the two candidates have similar properties, the presence of a scalar triplet in the Singlet + Triplet Scotogenic Model slightly changes the interaction of $\eta_R$ with the Higgs boson.
As a consequence, both its Higgs-mediated annihilation cross section as well as the $\eta_R$-nucleon interaction cross section contain a term dependent on $\mu_2$ and on $v_\Omega$
(see the relevant vertex in Appendix \ref{app:diagrams}).
This is nonetheless weighted by the (small) mixing between $h^0$ and $H$. As a result the $\eta_R$ dark matter phenomenology turns out to be very similar in both models. The real advantage of the Singlet + Triplet Scotogenic Model comes from the enlarged viable parameter space, especially at low $\eta_R$ masses, as it avoids the unwanted spontaneous breaking of the $\mathbb{Z} _{2}$ parity symmetry~\cite{Merle:2016scw}.

\subsection{Indirect detection}

\begin{figure}[!h]
\begin{center}
 \includegraphics[scale=0.55]{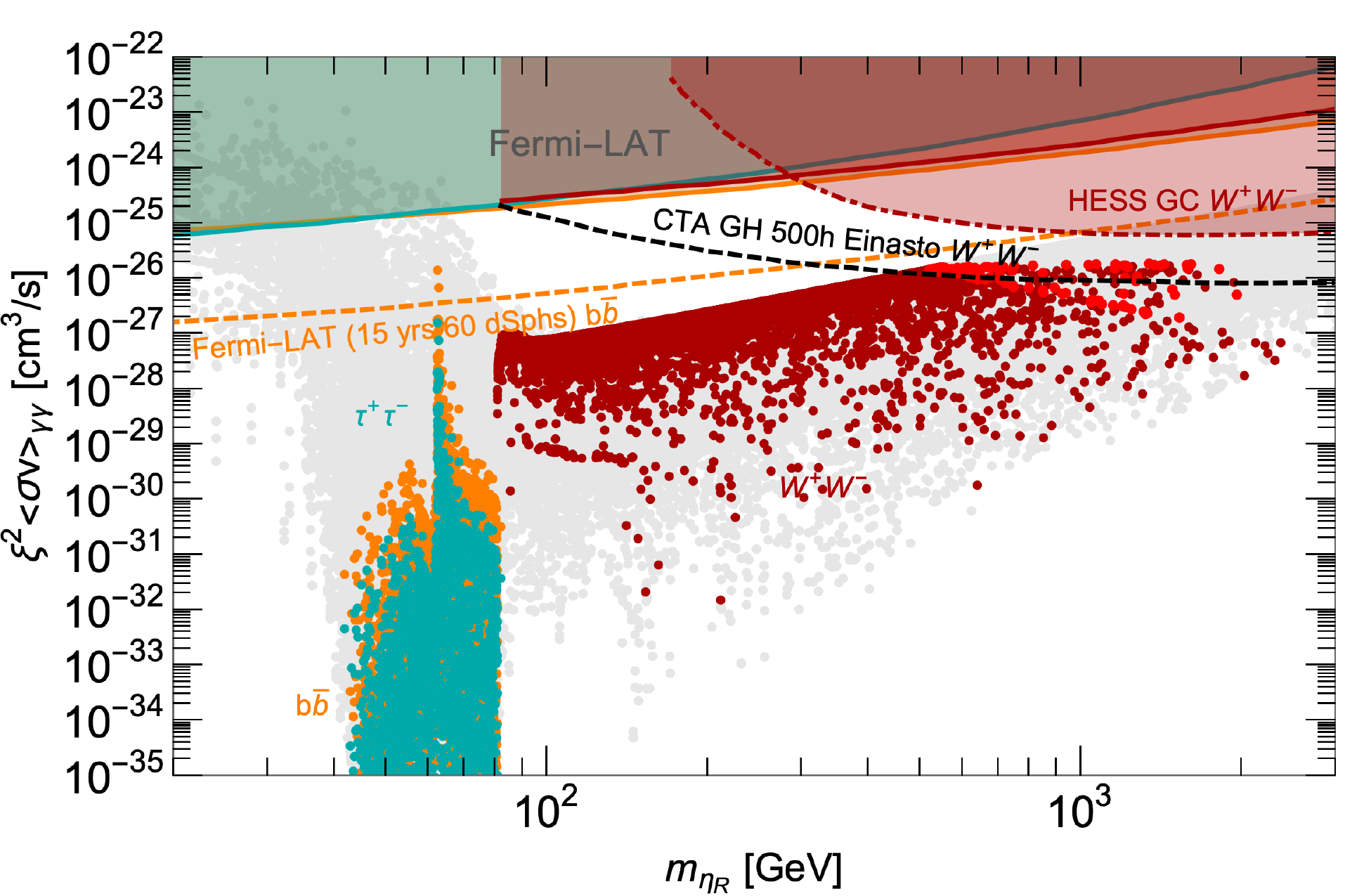}
\end{center}
\caption{Predicted $\eta_R$ annihilation cross section into $\gamma$ rays -- weighted by the relative abundance --  for annihilations to $b \bar{b}$ (orange), $\tau^+ \tau^-$ (dark cyan) and $W^+ W^-$ (dark and light red) final states. The orange, dark cyan and dark red plain lines refer to the corresponding 95\% C.L. upper limits currently set by the Fermi-LAT with $\gamma$-ray observations of dSphs~\cite{Ackermann:2015zua}. The dark red dot-dashed curve is the current upper limit obtained by H.E.S.S. using GC data~\cite{Abdallah:2016ygi}. We also compare with sensitivity projections for Fermi-LAT ($b \bar{b}$, 60 dSphs and 15 years of data)~\cite{Charles:2016pgz} and for CTA (GC, $W^+ W^-$)~\cite{Acharya:2017ttl}. See text for more details.}%
    \label{fig:DM_indirect}%
\end{figure}

If $\eta_R$ annihilates into SM products with a cross section near the thermal relic benchmark value, it may be detected indirectly. 
Among its annihilation products, $\gamma$ rays are probably the best messengers since they proceed almost unaffected during their propagation, thus carrying both spectral and spatial information.
First we consider prospects of detecting $\gamma$ rays from $\eta_R$ annihilations by considering the continuum spectrum up to the $\eta_R$ mass which originates from decays of the annihilation products.
We consider annihilations into $b \bar{b}$, $\tau^+ \tau^-$ and $W^+ W^-$ to compare with current limits set by the Fermi Large Area Telescope (LAT) satellite~\cite{Ackermann:2015zua} and HESS telescope~\cite{Abdallah:2016ygi}. 
We show in Fig.~\ref{fig:DM_indirect} the results of our numerical scan of the annihilation cross section (weighted by  $\xi^2$ and by the correspondent branching ratio) versus the $\eta_R$ mass, for $\eta_R$ annihilating into $b \bar{b}$ (orange points), $\tau^+ \tau^-$ (dark cyan) and $W^+ W^-$ (dark red).  
Grey points are excluded by any of the constraints listed in section~\ref{Ssec:Constraints}. Points in light red are solutions with relic abundance falling exactly within the $3\sigma$ band measured by Planck. 
In the same figure we also show the 95\% C.L. upper limits currently set by the Fermi-LAT with $\gamma$-ray observations of Milky Way dwarf spheroidal satellite galaxies (dSphs) based on 6 years of data processed with the Pass 8 event-level analysis~\cite{Ackermann:2015zua} (plain lines assuming annihilation into $b \bar{b}$ (orange), $\tau^+ \tau^-$ (dark cyan) and $W^+ W^-$ (dark red)). 
Moreover we show as a red dot-dashed curve the current upper limit obtained by H.E.S.S. using Galactic Center (GC) $\gamma$-ray data accumulated over 10 years~\cite{Abdallah:2016ygi}, assuming a $W^+ W^-$ channel and an Einasto dark matter density profile.  
Finally, we also depict sensitivity projections for Fermi-LAT from a stacked analysis of 60 dSphs and 15 years of data, in the $b \bar{b}$ channel~\cite{Charles:2016pgz}  (dashed orange) and for CTA, for the Milky way galactic halo target, $W^+ W^-$ channel and an Einasto dark matter density profile~\cite{Acharya:2017ttl}. 
Although current limits lie a couple of orders of magnitude above the predicted signals in this model, future data from Fermi-LAT and CTA offer promising prospects, eventually allowing one to test part of the parameter space both in the low ($\sim 70$ GeV) as well as in the high ($\gsim 500$ GeV) mass regions
~\footnote{ Note that for annihilations of non-relativistic $\eta_R$ occurring at the current epoch, the cross section and hence its indirect detection flux can be affected by a non-perturbative correction, the Sommerfeld enhancement~\cite{1931AnP...403..257S,Hisano:2003ec,Hisano:2004ds,ArkaniHamed:2008qn,Chowdhury:2016mtl}. This occurs when $m_{\eta_R} \gg  M_W (M_Z)$ and $\eta_R$ is almost degenerate in mass with $\eta^\pm (\eta_I)$.  The multiple exchange of $W$ ($Z$) bosons would induce a long range attractive force, thus leading to an enhancement of the annihilation cross section at low dark matter velocities, compared to its tree-level value.}. 

Besides $\gamma$ rays, charged cosmic rays can be used to probe $\eta_R$ as a dark matter candidate. 
The positron fraction measured by PAMELA~\cite{Adriani:2008zr,Adriani:2013uda} and more recently by AMS-02~\cite{Aguilar:2013qda,Bergstrom:2013jra}, allows us to place constraints on annihilating WIMPs, which are particularly stringent in the case of annihilations to the first two generations of charged leptons. 
In our scenario, light $\eta_R$ annihilate mainly to $\tau^+ \tau^-$, as can be seen from Fig.~\ref{fig:DMBR}.
As a result bounds from cosmic positrons are less relevant than those from $\gamma$ rays. 
In addition to cosmic-ray positrons, AMS-02 has also provided a high-precision measurement of the cosmic-ray antiproton spectrum~\cite{Aguilar:2016kjl}. 
These can be translated into upper limits on hadronic dark matter annihilation, which can be a factor of few stronger than those from $\gamma$-ray observations of dSphs~\cite{Cuoco:2016eej,Cholis:2019ejx}.
Since these results rely on a careful treatment of systematic uncertainties, namely the antiproton production cross-section, and the modelling of the effect of solar modulation we decided not to include them here and leave it for a dedicated work. 
Similarly, searches for anti-deuterium or anti-helium events could potentially provide a powerful probe of  $\eta_R$ annihilations~\cite{Donato:1999gy,Carlson:2014ssa,Cirelli:2014qia,Coogan:2017pwt}, although also affected by substantial uncertainties.

\section{Scalar dark matter signatures at the LHC}
\label{sec:lhcDM}

In this section we confront our scalar dark matter candidate with the latest data from particle colliders, in particular from the LHC run at $\sqrt{s}=13$ TeV.
As in any model with a dark matter candidate, the generic signature to be searched for is missing energy ($\met$), measured from the total transverse momentum recoil of the visible particles in the event (see for instance \cite{ATLAS:2018ghb, CMS:2016ljj}). 

\begin{figure}[!h]
\centering
\begin{minipage}{0.25\textwidth}
   \includegraphics[width=\textwidth]{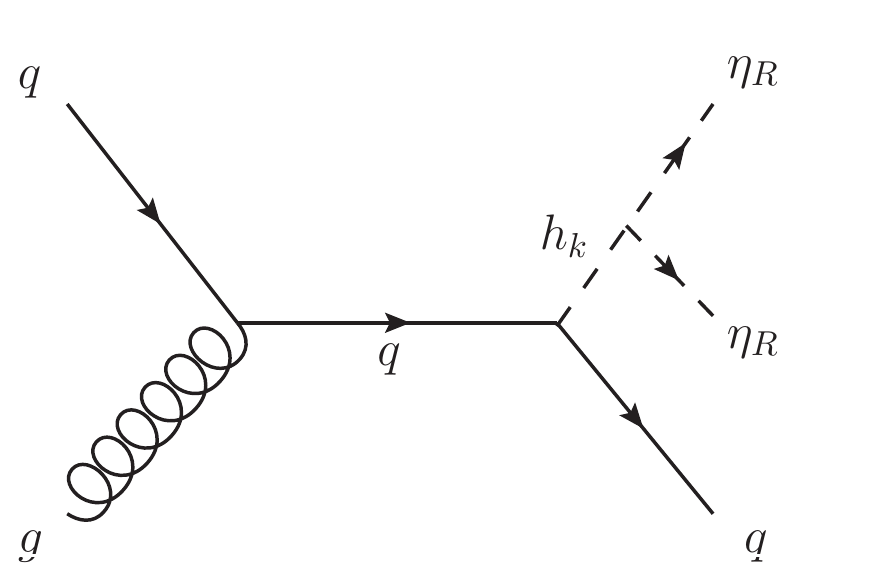}
\end{minipage}
\begin{minipage}{0.33\textwidth}
    \includegraphics[width=\textwidth]{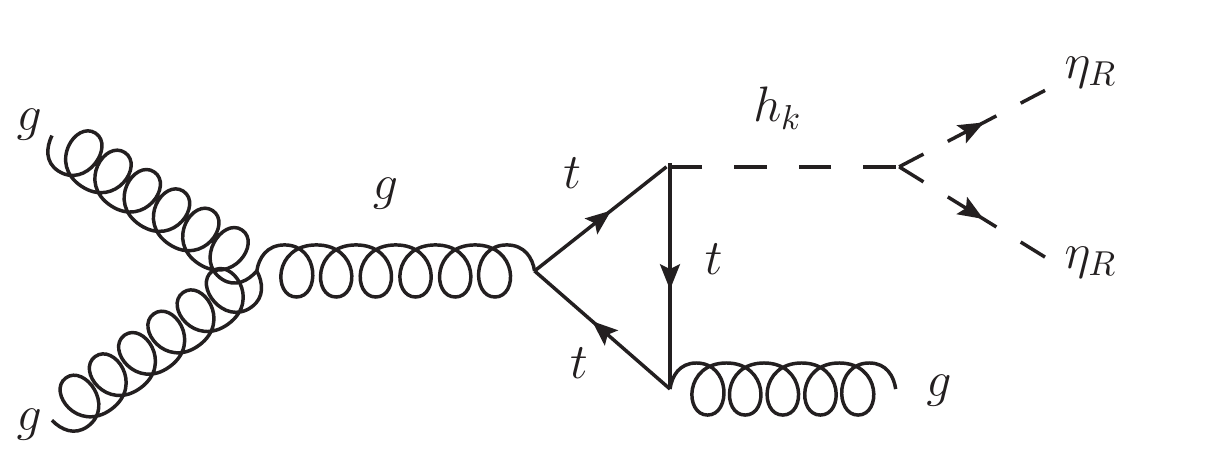}
\end{minipage}
\begin{minipage}{0.3\textwidth}
    \includegraphics[width=\textwidth]{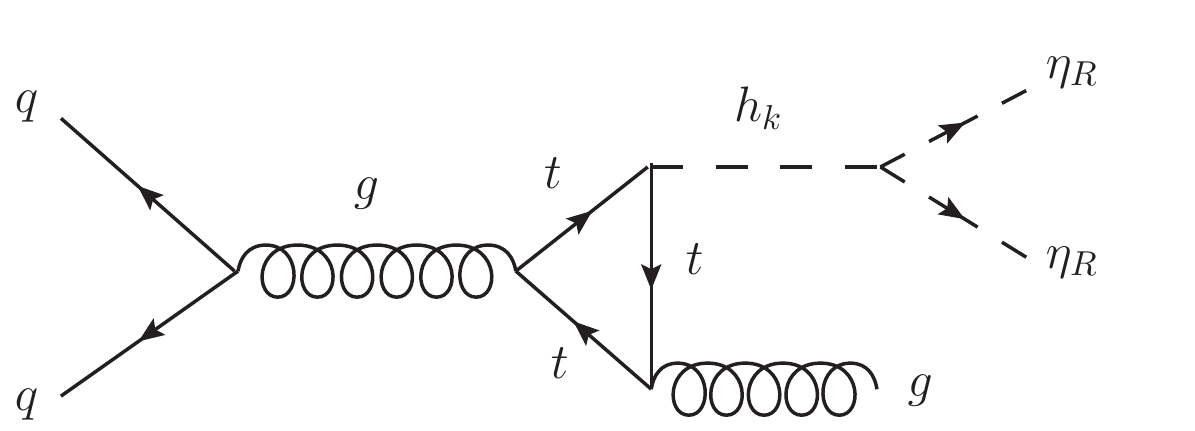}
\end{minipage}
\end{figure}
\begin{figure}[H]
\centering
\begin{minipage}{0.29\textwidth}
    \includegraphics[width=\textwidth]{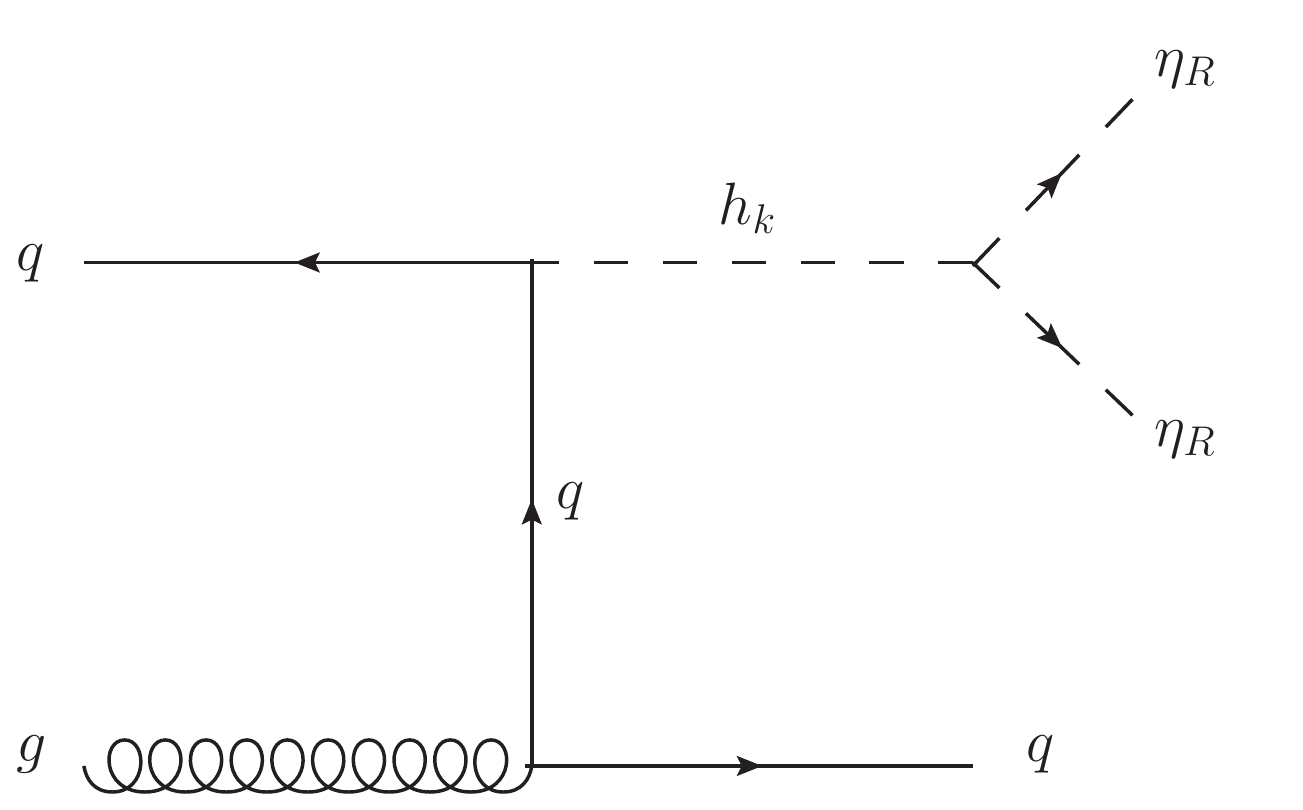}
\end{minipage}
\begin{minipage}{0.29\textwidth}
    \includegraphics[width=\textwidth]{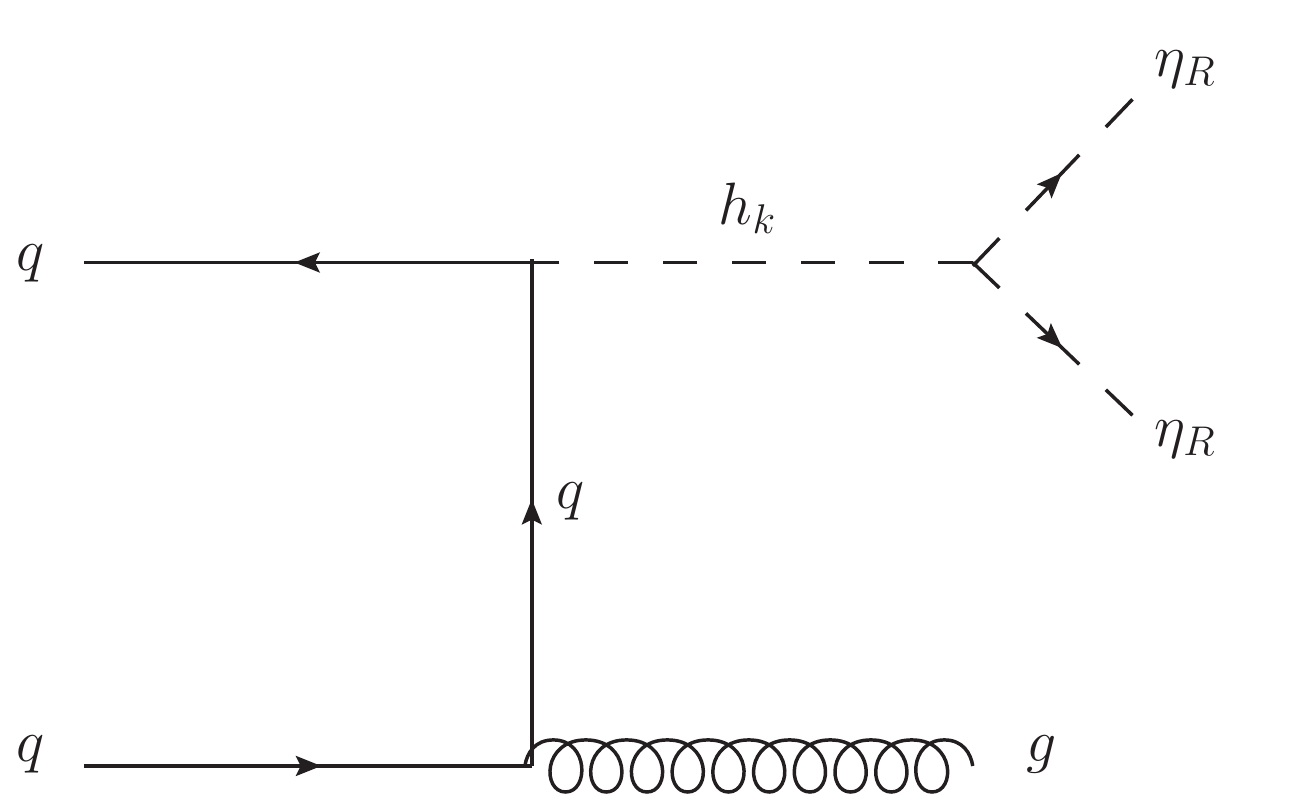}
\end{minipage}
\begin{minipage}{0.29\textwidth}
    \includegraphics[width=\textwidth]{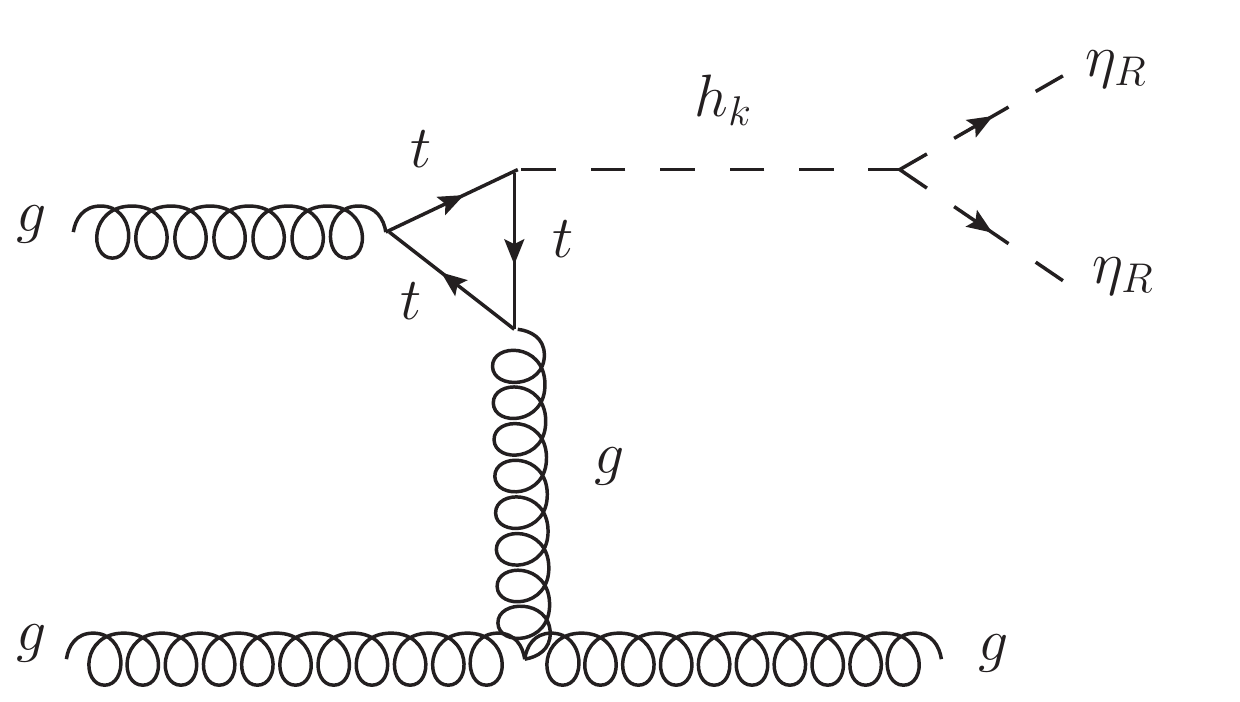}
\end{minipage}
\caption{Relevant Feynman diagrams for mono-jet production through $\eta_{R} \eta_{R}+j$ at the LHC; here $h_k\equiv$ $h^{0}$ or $H$.}
\label{fig:monojetRR} 
\end{figure}

In the Triplet + Singlet Scotogenic Model typical signatures are  $\met + X$, where $X$ can be one or two jets~\cite{Aaboud:2016tnv, Haisch:2018hbm}, two leptons~\cite{Aaboud:2018ujj} or one photon \cite{Aaboud:2017dor}. 
Although all of them are in principle interesting, we have checked numerically that in our scenario the most promising one is $\met +$ jet (mono-jet). 
In the following we will focus on mono-jet final states, arising from $p p \to \eta_{R} \eta_{R}+g$ and $p p \to \eta_{R} \eta_{R}+q$ processes. 
Here ones looks for events with one high-$p_T$ jet (higher than $100-200$ GeV in the central region of the detector, with pseudorapidity $|\eta|<2.4$) and $\met$ above roughly $200$ GeV in the $13$ TeV analyses for the ATLAS and CMS detectors~\cite{Aaboud:2016tnv, CMS}. 
The dominant irreducible SM background for this channel comes from $Z + j$, with the $Z$
  boson subsequently decaying invisibly $Z \to \nu \bar{\nu}$. There is also a subdominant irreducible background from $W + j$, with $W \to \tau \nu$, where the $\tau$ decays hadronically.
  In addition, there are backgrounds from $W + j$ with $W \to \mu \nu$ or $e \nu$, where the lepton is either missed or misidentified as a jet.
  However, $Z + j$ constitutes approximately 60\% of events.\\

\begin{figure}[!htb]
\centering
  \begin{minipage}{0.26\textwidth}
    \includegraphics[width=\textwidth]{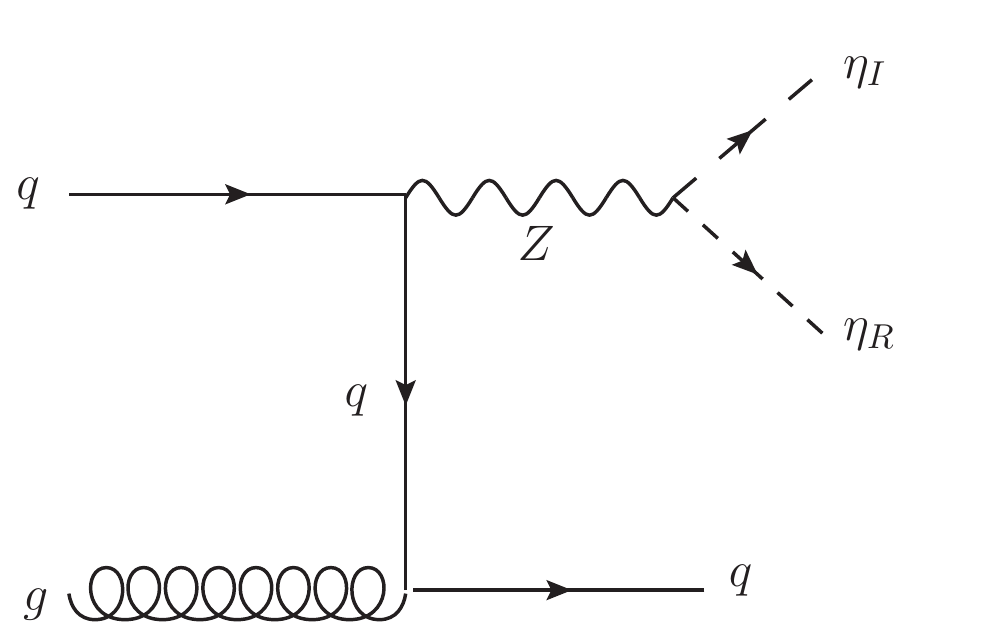}
  \end{minipage}
 \begin{minipage}{0.26\textwidth}
    \includegraphics[width=\textwidth]{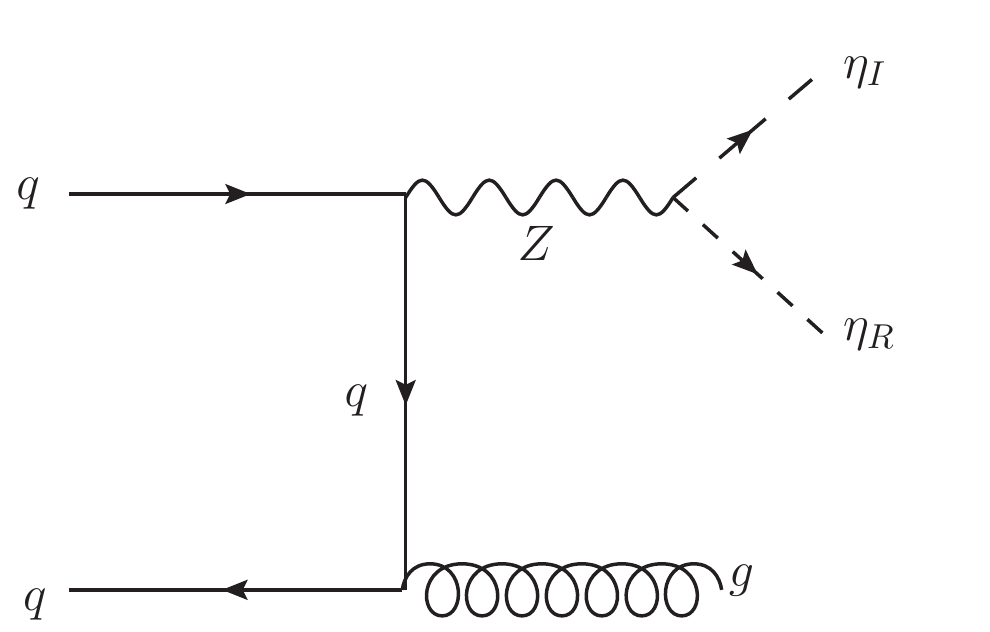}
    \end{minipage}
  \begin{minipage}{0.26\textwidth}
    \includegraphics[width=\textwidth]{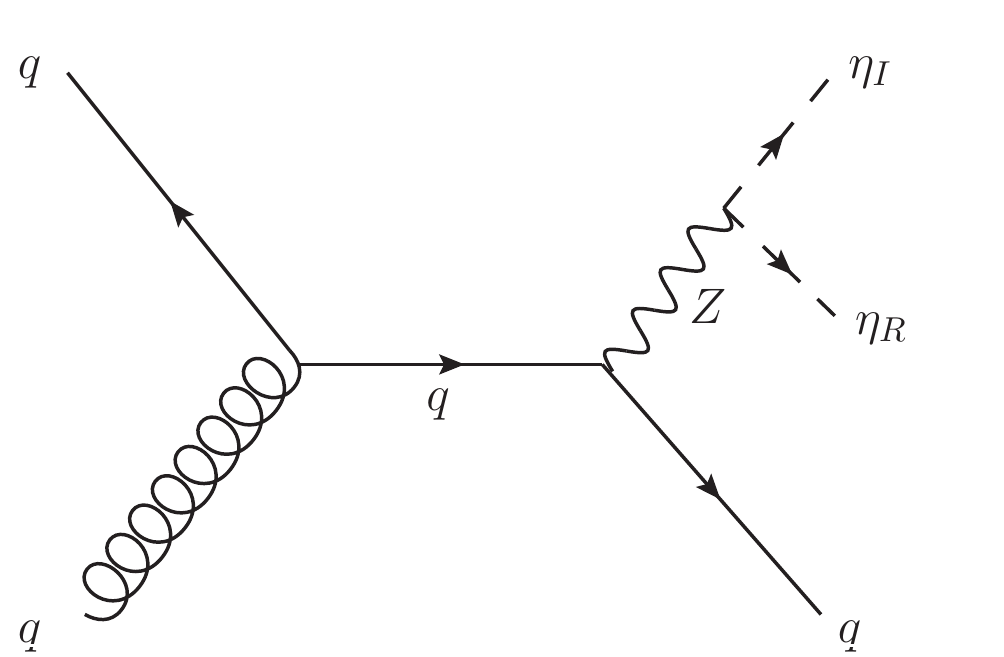}
\end{minipage}
\caption{Feynman diagrams illustrating $Z$-mediated production of $\eta_{R} \eta_{I}+j$ at the LHC.}
\label{fig:monojetRI}
\end{figure}

At leading order, the relevant Higgs-mediated Feynman diagrams for mono-jet events are shown in Fig. \ref{fig:monojetRR}. 
In all cases, the dark matter is produced via the decay of a neutral scalar ($h^0$ or $H$), produced from its interaction with quarks, or through its effective coupling to gluons. The latter involves a top quark loop and enters in gluon-gluon fusion (ggF) processes. 
An important point is that in ggF processes only the SM-like Higgs doublet couples with fermions. Indeed, since $H$ is mainly a triplet, its coupling with quarks is suppressed. 
The interaction vertex between $\eta_{R}$ and $h_{k}$ is given in Appendix~\ref{app:diagrams}. 
Note that if the mass difference between $\eta_I$ and $\eta_R$ is small, $\eta_I$ should also contribute to the invisible final states.
In this case, $\eta_{I}$ would subsequently decay to $\eta_{R}$ plus soft fermions or jets which are not energetic enough to be detected. 
Besides Higgs mediation, the mono-jet signal can proceed also via Z-mediation. Therefore we also include the contributions shown in Fig. \ref{fig:monojetRI}, which are described as
  $p p \to \eta_{R} \eta_{I}+g$ and $p p \to \eta_{R} \eta_{I}+q$ processes. 
Finally, we must mention that in this same scenario of small mass differences, a pair of $\eta_I$ can also be produced. 

\subsection{Benchmark Points}
  
  The constraints previously described in Section~\ref{Ssec:Constraints} restrict the parameter space allowed by a vast array of experimental probes, among which are the relic density, direct detection and indirect detection analyses.
  Motivated by these preliminary studies, we now investigate using the {\tt CheckMATE 2} collaboration tools \cite{Dercks:2016npn, Sjostrand:2007gs, deFavereau:2013fsa, Cacciari:2011ma, Read:2002hq} whether the solutions that satisfy all experimental limits in section~\ref{Ssec:Constraints} could lead to detectable dark matter mono-jet signals at LHC $13$ TeV.
This code allows us to determine whether or not a given parameter configuration of our model is excluded at  95\% C.L, recasting the results of the simulated model in terms of the existing analyses of the LHC, which automatically include the simulation and elimination of the background.
Indeed, for each signal region, {\tt CheckMATE 2} computes the expected number of signal events $S$ after cuts, 
and directly compare it to the $95$\% C.L. upper limit $S^{95}_{exp}$, given a signal error $\Delta S$. 
The most relevant analysis for our study is Ref.~\cite{Aaboud:2016tnv}.

In this way, we identify two interesting benchmark points which survive the entire set of constraints described in Section \ref{Ssec:Constraints} (including the theoretical ones,
    such as the conservation of the $\mathbb{Z} _{2}$ parity symmetry up to higher energy scales) and shown in Table~\ref{tab:b1}.
  Values of the relevant parameters and the corresponding scalar spectrum are summarised. We also show in this table the value of observables obtained in Section \ref{Ssec:Constraints} for each benchmark.
  The main difference between the two benchmark points is the value of $H$ mass, which is governed by $\mu_{2}$ and $v_{\Omega}$.
  However, because this heavy scalar is mainly triplet, its coupling with quarks in the ggF processes is suppressed, so that a significant change in its mass is not expected to lead to a large variation in the magnitude of the cross sections. 

\begin{center}
 \begin{table}[htb]
\scalefont{1}
\centering
\vspace{0.7cm}
\begin{tabular}{|c|| c c c| }
\hline
{\bf Parameters} & {\bf Benchmark 1} &{\bf Benchmark 2} &{\bf Units}  \\ 
\hline \hline                              
$\lambda_{3}$ &$3.64\times10^{-5}$ &$-1.64\times10^{-5}$ & - \\
$\lambda_{4}$ &$7.02\times10^{-7}$ &$-3.29\times10^{-7} $ &-\\
$\lambda_{5}$ &$-1.8\times10^{-2}$ &$-1.45\times10^{-2}$&-\\
$\lambda_{\eta}^{\Omega}$ &$-1.32\times10^{-5}$ &$-7.11\times10^{-6}$   & \\
$\mu_{2}$  &$-4.57\times10^{-8}$ &$-1.59\times10^{-1}$ & GeV\\
$v_{\Omega}$  &$2.43\times10^{-4}$ &$9.21\times10^{-1}$ & GeV\\
$m_{\eta}^2$ &$3678.17$ &$2851.39$ & GeV$^2$ \\
\hline\hline 
{\bf Scalar masses}  & & & \\      
\hline\hline                      
$m_{\eta_{R}}$  &$55.92$ &$ 49.09$ & GeV\\
$m_{\eta_{I}}$ &$65.04$ &$57.38$ & GeV\\
$m_{h^{0}}$  &$124.68$ &$125.54$ & GeV\\
$m_{H}$ &$425.9$ &$834.45$ & GeV\\ 
\hline\hline
{\bf Constraints} & &  &\\
\hline\hline 
$\Omega h^2$ &$0.0107$ &$0.0129$ &-\\
BR$(h^0\rightarrow inv.)$ & $0.155489$ &$0.12939$ &-\\
BR$(\mu\rightarrow e\gamma)$ &$7.33\times10^{-29}$ &$8.55\times10^{-32}$ &-\\
BR$(\mu\rightarrow eee)$ &$3.75\times10^{-30}$ &$1.01\times10^{-30}$ &-\\
CR$(\mu^{-}, Au\rightarrow e^{-}, Au)$ &$3.88\times10^{-29}$ &$1.40\times10^{-29}$ &-\\
BR$(h^0\rightarrow\gamma\gamma)$ &$0.00226748$ &$0.00212008$ &-\\
$\Delta a_{\mu}$ &$2.18\times10^{-14}$ &$2.15\times10^{-14}$ &-\\
$\sigma_{SI}$ &$5.953\times10^{-10}$ & $4.862\times10^{-10}$ &pb \\
\hline
\end{tabular}
\caption{Benchmark points which survive the entire set of constraints described in Section \ref{Ssec:Constraints} and corresponding parameters relevant to the calculation of diagrams in $\met+$jet final states.}
\label{tab:b1}
\end{table}
\end{center}

\subsection{Mono-jet signatures at the LHC $\sqrt{s}={13}$ TeV}\label{sec:13TeV}

We display in Tab.~\ref{tab:check1} the {\tt CheckMATE 2} results for the evaluation in the $\met+$jet channel (corresponding to an integrated luminosity of $36.1 {\rm fb}^{-1}$ in the $\sqrt{s}=13$ TeV analysis) for the two benchmark points of Tab~\ref{tab:b1}. For this study, the cross sections shown in Tab.~\ref{tab:check1} correspond to both contributions to the final state studied: $Z$ boson (Fig. \ref{fig:monojetRI}) and Higgs-mediated processes (Fig. \ref{fig:monojetRR}), respectively. 

\begin{table}[!htb]
\centering
 \scalefont{1.0}
\vspace{0.5cm}
\begin{tabular}{|c|c|c|}
\hline
{\bf Quantity} &{\bf Benchmark 1} &{\bf Benchmark 2}\\ 
\hline
{ $\sigma \pm d\sigma$ [fb]} &$787.791$ & $1074.62$  \\
{\bf $S\pm dS$} &$163.241\pm6.814$ &$421.3\pm12.784$ \\
{\bf $r$} &$0.220$ &$0.263$ \\ \hline
\end{tabular}
\caption{Results obtained with {\tt CheckMATE 2} based on the {\tt atlas\_conf\_2017\_060}~\cite{Aaboud:2016tnv} analysis by the ATLAS collaboration, for LHC data at $\sqrt{s}=13$ TeV.}
\label{tab:check1}
\end{table}

The main result of Tab.~\ref{tab:check1} is the value of the parameter $r$
 \begin{equation}\label{r}
 r \equiv \frac{S-1.96\Delta S}{S^{95}_{exp}}
\end{equation} 
calculated by {\tt CheckMATE 2}\footnote{According to algorithm definitions and taking into account experimental errors, a point in parameter space is considered excluded if the ratio $r\geq1.5$. If $r \leq 0.67$, the point is classified as compatible with the experimental results and is kept. Points with $0.67 < r < 1.5$  are regarded as ``potentially excluded'' in view of the systematic and theoretical errors. For more details see \cite{Domingo:2018ykx}.},
which translates into a significant number of signal events after the cuts, $S$.
These specific cuts are implemented by the ATLAS analysis in order to map out the associated regions of consistent parameter choices, and will be described later. 

Our dark matter candidate $\eta_R$ with mass around $\sim 50-60$ GeV and chosen to satisfy all theoretical and experimental constraints of Sec.~\ref{Ssec:Constraints} 
  would lead to a signature in the $\met+$jet channel in the ATLAS experiment.
  For that we require, for both benchmark points, that the leading jet has $p_{T}> 250$ GeV and $|\eta|<2.4$, separation in the azimuthal plane of $\Delta \phi (\text{jet}, p_{T}^{miss})>0.4$ between the missing transverse momentum direction and each selected jet.
  The difference between our benchmarks are the $\met$ thresholds. While for Benchmark 2 a $\met$ minimum of $500$ GeV is required, in the other case we take $\met>600$ GeV.   \\ [-.2cm]

For larger $\eta_{R}$ masses we investigate the behaviour of the cross sections at $\sqrt{s}=13$ TeV and the projected signal events at $\sqrt{s} = 14$ TeV. 
We assume the coupling $|\lambda_{345}|$ to lie in the range $[0.02, 0.9]$ and we fix the other parameters according to Benchmark 1 in Tab. \ref{tab:b1}.
We analyse $\eta_R \eta_R +j$ and $\eta_R \eta_I +j$ separately because the rate of these processes depends on different parameters
and we want to analyze their contributions to the total cross section separately.\\ [-.2cm]

\begin{figure}[!htb]
\centering
\includegraphics[width=0.47\linewidth]{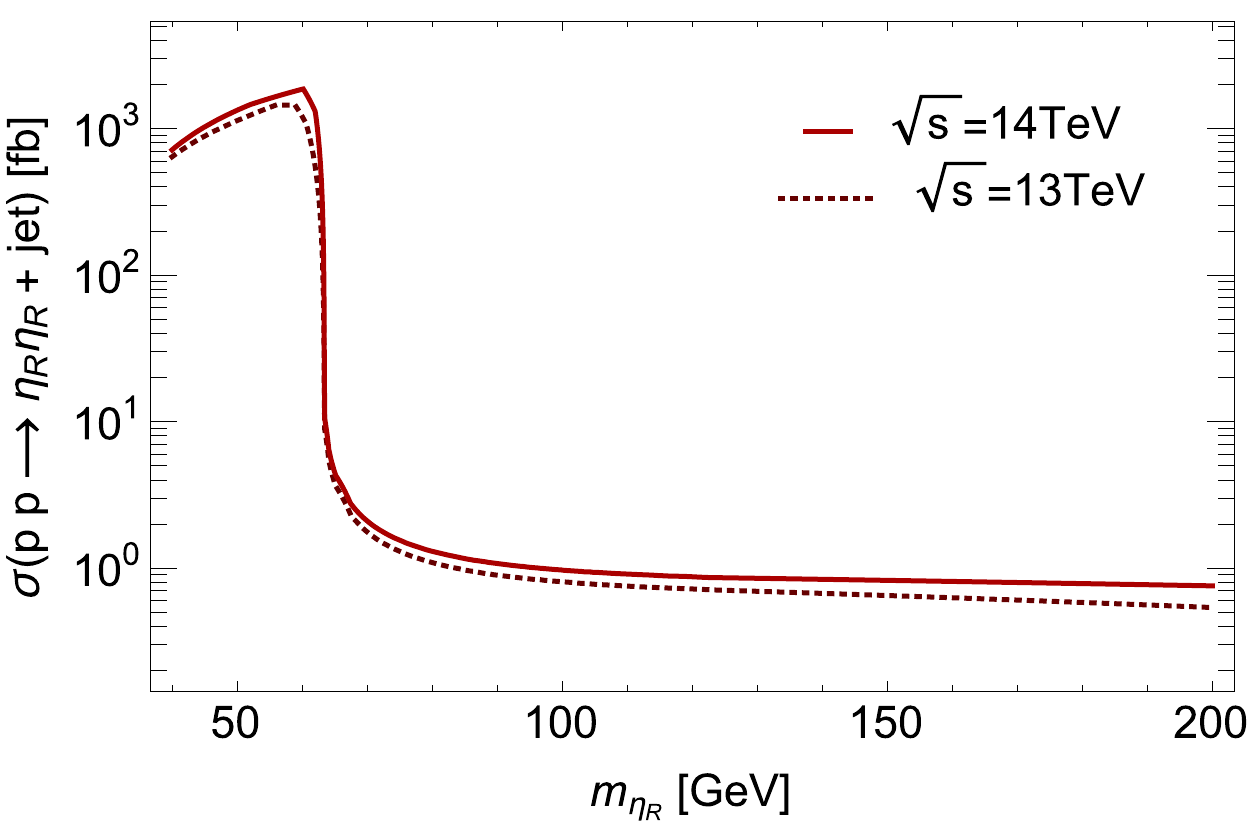}\hspace{0.75cm}
 \includegraphics[width=0.47\linewidth]{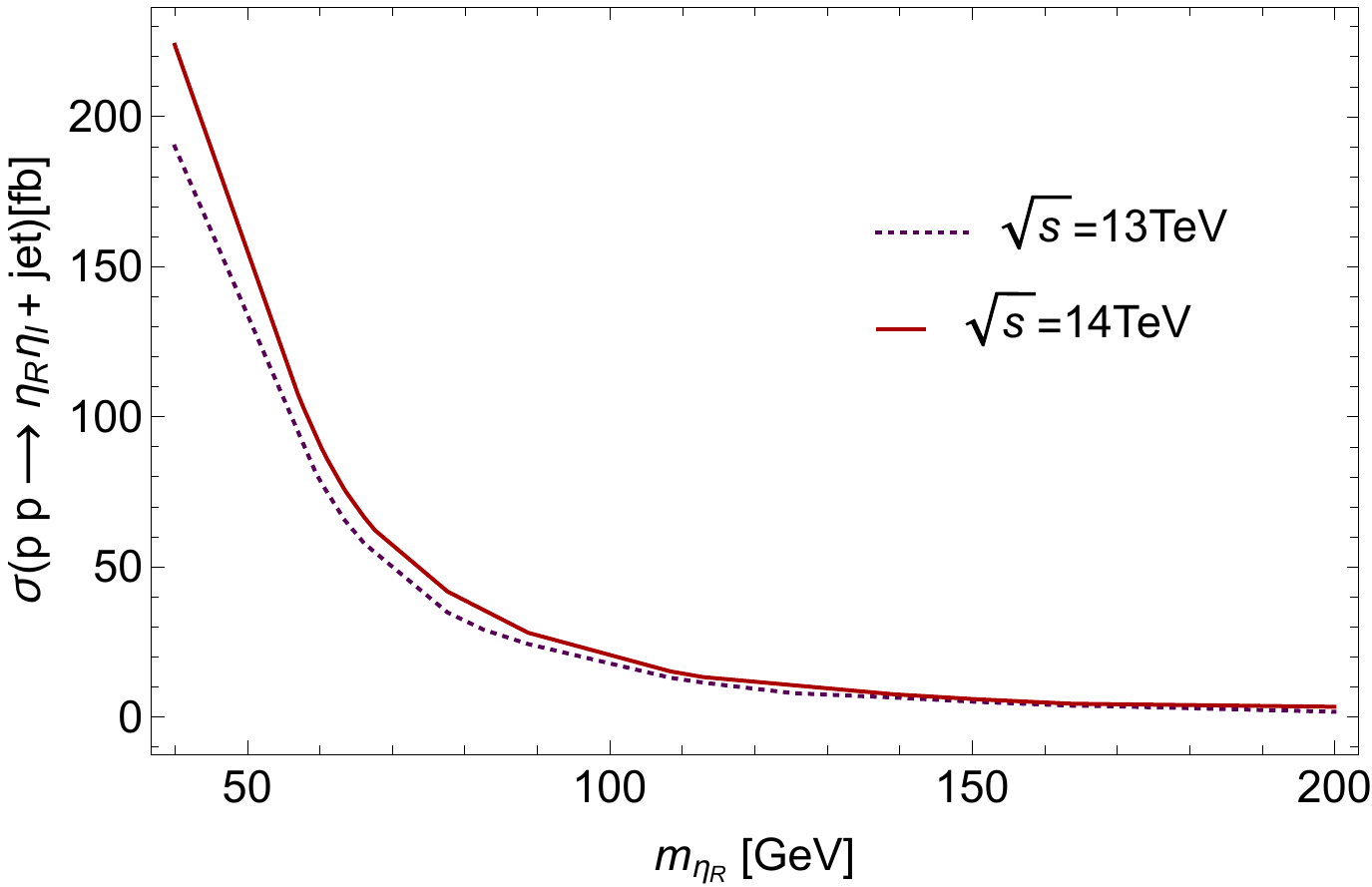}
 \caption{Cross sections of mono-jet signals at LHC $\sqrt{s}={13~(14)}$ TeV. The left panel shows the Higgs boson mediated events from $p p \to \eta_{R} \eta_{R}+$ jet.
    The maximum value of the cross section is $\sim1400(1800)$ fb for $\sqrt{s}=13~(14)$ TeV respectively. 
The right panel is the Z-mediated process, $p p \to \eta_{R} \eta_{I}+$ jet, with peak contribution $\sim190~(220)$ fb.}
\label{fig:monojet}
\end{figure}

  In Fig. \ref{fig:monojet} we present the production cross section for $\met+$ jet process at LHC $\sqrt{s}=13~(14)$ TeV.
Using {\tt Madgraph5} \cite{Alwall:2014hca} we simulate events with an initial cut of $p_{T}^{jet}>100$ GeV, according to the latest analyses in mono-jet searches~\cite{Belyaev:2018ext, Belyaev:2016lok}.
Since the relevant processes leading to these events are mediated by mainly the SM Higgs (left panel) and $Z$ boson (right panel), one has the characteristic peaks at $m_ {\eta_{R}}\sim m_{h^0}/2$
  and at $m_ {\eta_{R}}\sim m_{Z}/2$ respectively, providing larger cross sections in these mass ranges.
Therefore, the Higgs boson mediated processes are dominant up to $m_{\eta_R}\sim 60$ GeV and also contribute in the range $\sim [700-1400]$ fb ($13$ TeV).
  In addition, $Z$-mediated processes complements the search for $p p \to \eta_{R} \eta_{R} +$ jet process at the LHC.
For this mass range, the cross sections are $\sim [190-80]$ fb while,
for dark matter masses between $[65-200]$ GeV, we have $\sim [70-5]$ fb,
providing a sizeable contribution to the total mono-jet cross section, which could be within LHC sensitivity.
At $\sqrt{s} = 14$ TeV the cross section increases by a few fb.
These results agree with expectations of other models, such as the Inert Higgs Doublet Model, whose contributions to this signal are very similar \cite{Belyaev:2018ext}.
In summary, one sees that there are good prospects for probing the mono-jet signal at the LHC for dark matter masses up to $\sim60$ GeV.

\begin{figure}[!hbt]
\centering
  \includegraphics[width=.6\linewidth]{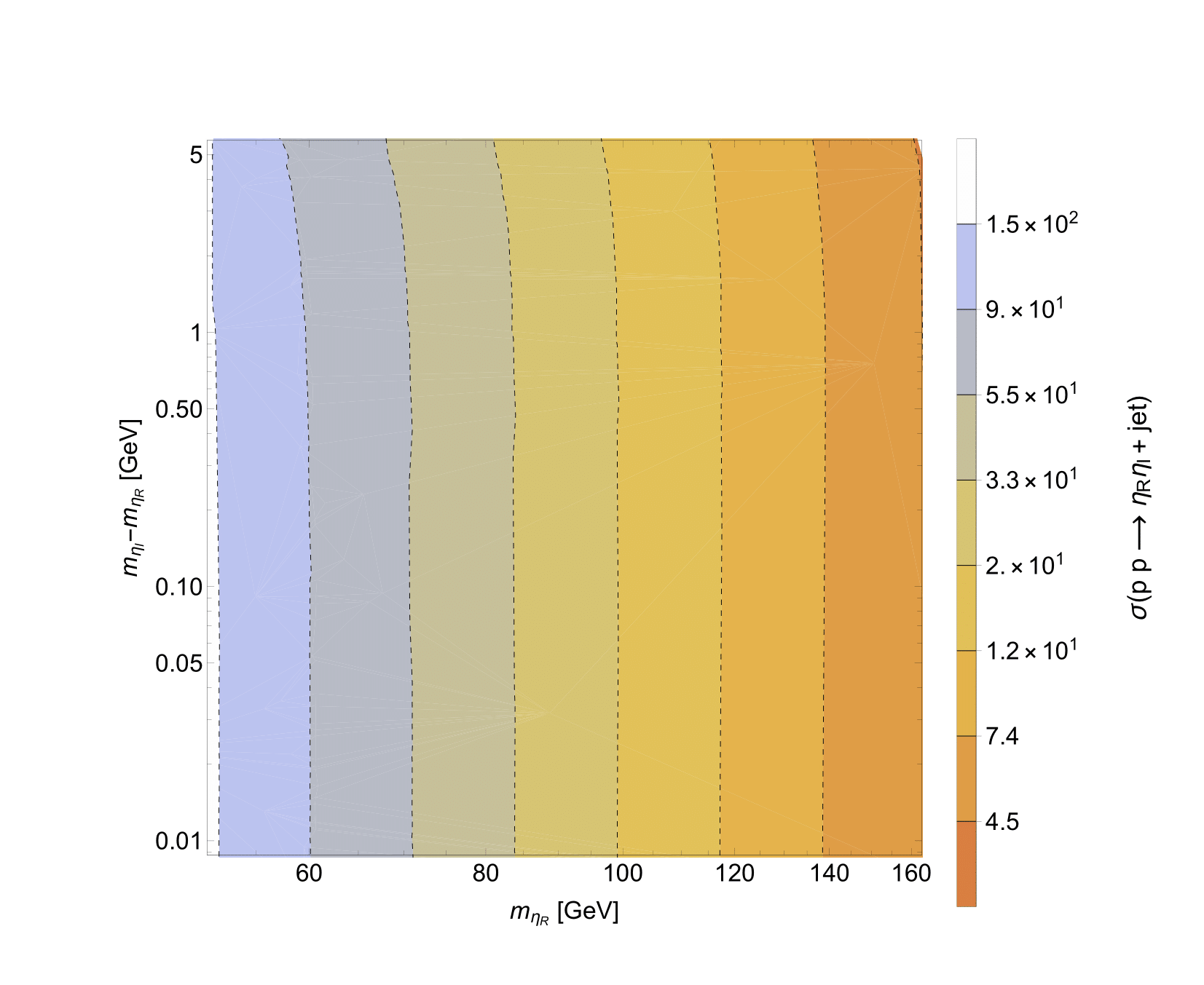}
  \caption{Mass difference $m_{\eta_{I}} - m_{\eta_{R}}$ as a function of $m_{\eta_{R}}$ in mono-jet events mediated by the Z boson, $p p \to \eta_{R} \eta_{I}+$ jet.
    The color shades represent values of the cross section in fb.}
\label{fig:Z}
\end{figure}

There are regions of parameters in which $\eta_I $ and $\eta_R$ are relatively close in mass, as shown in Figure \ref{fig:Z}.
  This as required for model consistency, as the mass difference between these particles is intimately connected with the smallness of neutrino mass
  as generated in the scotogenic picture. This requires the violation of lepton number through the value of $\lambda_5$, as seen by Eqs. \ref{eta0} and \ref{eq:metaRI}. 
Indeed, if $m_{\eta_R}-m_{\eta_I}$ is small we can obtain neutrino mass square differences, as needed to account for neutrino oscillation data \cite{deSalas:2018bym}.
Moreover, the particles produced from the decay $\eta_I \to \eta_R$+X are not energetic enough to have the trajectories reconstructed by the detector (soft particles),
leading to our $\met +$ jet final state signal.\\[-.2cm]

As already commented in previous sections, as a result of its small coupling with quarks, the heavy neutral scalar $H$ does not influence significantly our signal.
  As shown in Ref. \cite{Diaz:2016udz}, the production cross section of $H$ at the LHC is $3$ to $5$ orders of magnitude smaller than the production of
  the SM Higgs boson, independent of the center-of-mass energy.
  Hence our results for the scalar dark matter jet + missing energy final states within the Singlet + Triplet Scotogenic Model should also hold within the simplest
  Scotogenic scenario of~\cite{Ma:2006km}.

\section{Summary and conclusions}
\label{sec:Conclusions}

In this work we have reexamined the generalized version of the minimal Singlet + Triplet Scotogenic Model, in which dark matter emerges naturally as the mediator of neutrino mass generation
and its stability follows from the same $\mathbb{Z} _{2}$ symmetry also responsible for the radiative origin of neutrino masses.
Notice that, while the simplest model of Ma~\cite{Ma:2006km} fails to be consistent over a wide range of parameters~\cite{Merle:2015ica},
  our generalized scotogenic model is the minimal one allowing for a conserved $\mathbb{Z} _{2}$ symmetry all the way up to high mass scales~\cite{Merle:2016scw}.
We have assumed dark matter to be a scalar WIMP and we have presented a full numerical analysis of the signatures expected at dark matter detectors as well as collider experiments.
{We have shown that direct detection data from XENON-1T already disfavour part of the parameter space, in particular solutions with mass in the $\sim 100$ GeV range.
    We have highlighted the importance of complementary searches, for instance via indirect detection with $\gamma$ rays.
    Although current limits from Fermi-LAT and HESS lie a couple of orders of magnitude above the predicted signals in this model, future sensitivities also for CTA offer promising prospects,
    eventually allowing to probe both the low and the high WIMP mass regions.}
We have identified the regions of parameters where dark matter predictions are in agreement with theoretical and experimental constraints,
such as those coming from neutrino oscillation data, Higgs data, dark matter relic abundance and direct detection searches.
We have also presented expectations for near future direct and indirect detection experiments. These will further probe the parameter space of our scenario.
 Finally, we have examined the collider signatures associated to the mono-jet channel at the LHC. 
In particular, we have found a viable light dark matter mass range in the region $50-60$ GeV.
This should encourage future studies at the upcoming high-luminosity run of the LHC.


\appendix 
\section{Feynman diagrams for relic abundance and direct detection searches}
\label{app:relic_diagrams}

\begin{figure}[!htb]
\centering
\includegraphics[scale=0.39]{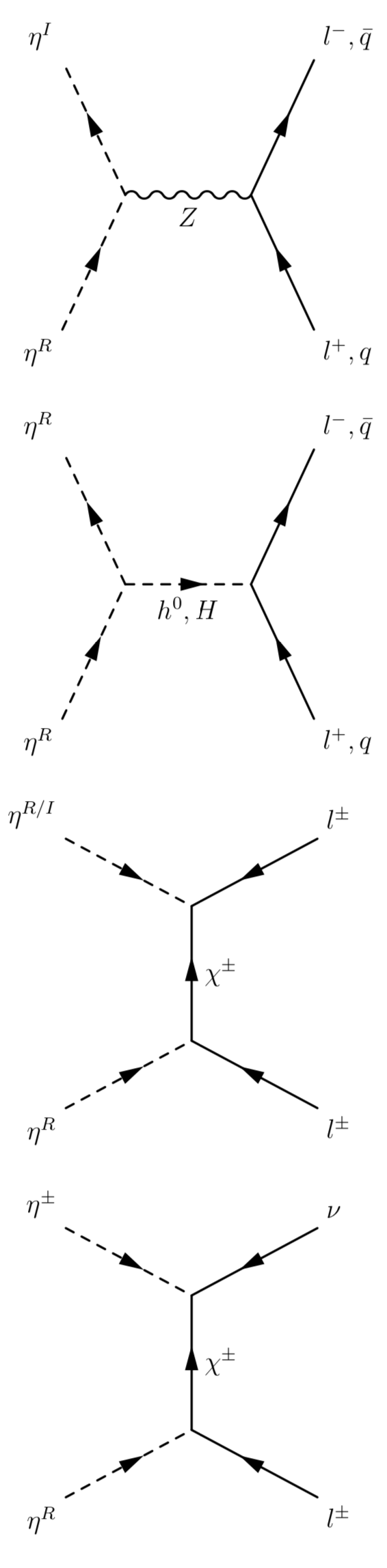}
\includegraphics[scale=0.39]{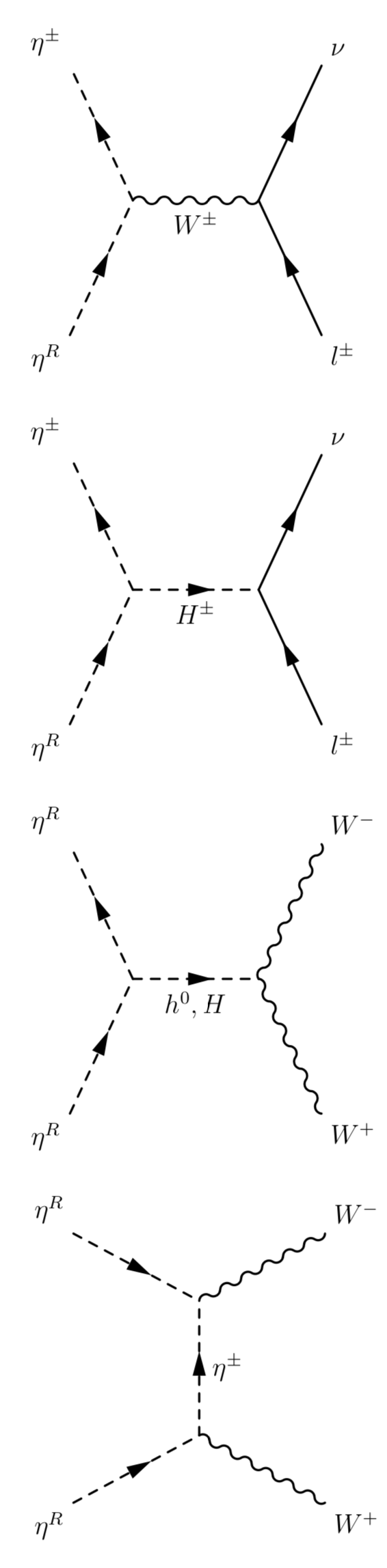}
\includegraphics[scale=0.39]{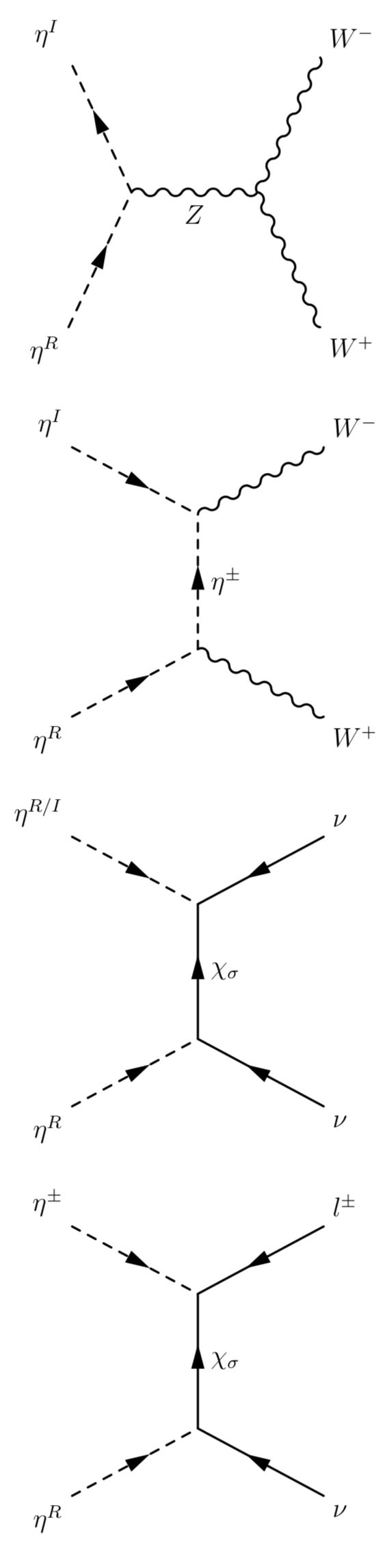}
\includegraphics[scale=0.39]{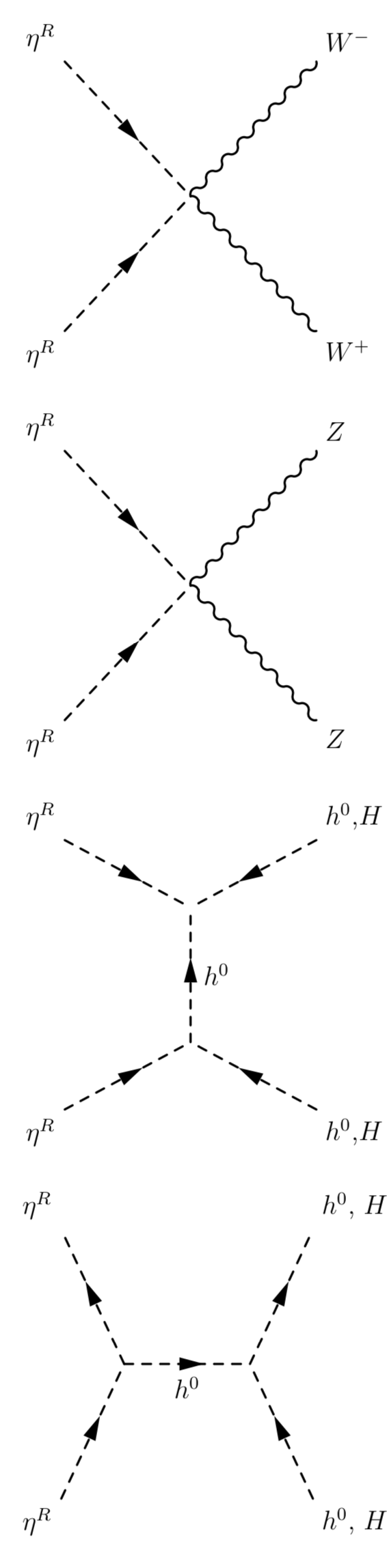}
\caption{Relevant annihilation and coannihilation diagrams contributing to the relic abundance of $\eta_R$.}
\label{fig:relic_diagrams}
\end{figure}

Here we present some of the main Feynman diagrams relevant to determine the cosmological relic density, assuming that $\eta_R$ is the dark matter. 
Fig.~\ref{fig:relic_diagrams} shows the main dark matter annihilation and coannihilation channels.
Besides the standard s-wave annihilation into quarks and gauge bosons, mediated by the SM-like Higgs boson, coannihilations with both  $\eta_R$ and  $\eta^\pm$ are possible. 
These can be mediated either by the $Z^0$ boson, or also by the new fermions $\chi_\sigma$.
These channels can lead to both charged or neutral leptons in the final state, and involve the contribution of the new Yukawas described in Section~\ref{sec:model}. 
Notice that these processes are not present in the simplest scotogenic constructions~\cite{Ma:2006km} nor in the case of the Inert Higgs Doublet Model~\cite{Diaz:2016udz}.
Diagrams with quartic interactions will appear when kinematically allowed, starting at $m_{\eta_R} \gtrsim 80$ GeV. 

\begin{figure}[h!]
\centering
\includegraphics[scale=0.3]{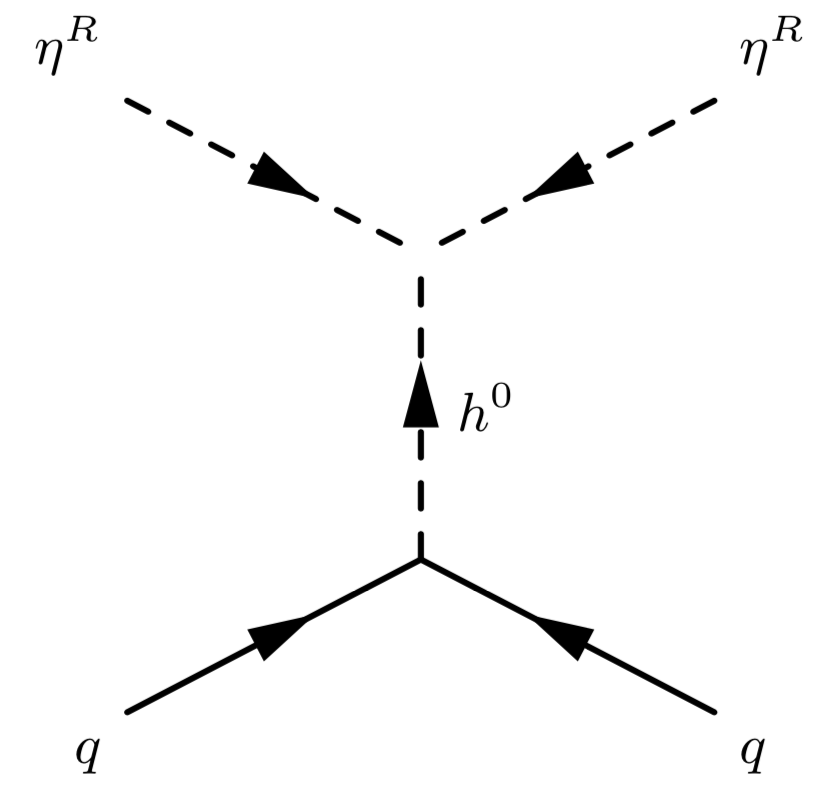}
\includegraphics[scale=0.3]{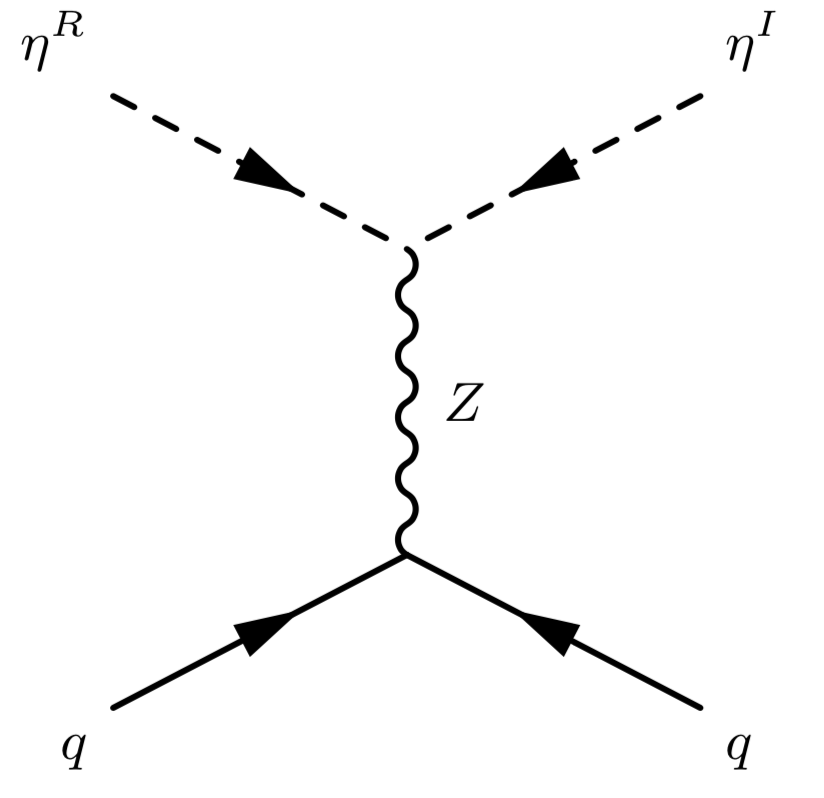}
\caption{Tree-level diagrams contributing to the elastic scattering of $\eta_R$ off nuclei via the Higgs exchange (left) and $Z^0$-boson exchange (right).}
\label{fig:DD_diagrams}
\end{figure}

The diagrams in Fig.~\ref{fig:DD_diagrams} contribute to the spin-independent $\eta_R$-nucleon elastic scattering cross section at tree level, discussed in Section~\ref{sec:DD}. 
The contribution of the diagram on the right is important only when the splitting between the masses of $\eta_R$ and $\eta_I$ is small (small $\lambda_5$ values) and leads to inelastic signals.
\label{app:DD_diagrams}

\section{Relevant Feynman rules for the Singlet+Triplet Scotogenic Model}
\label{app:diagrams}

\begin{figure}[h!]
\begin{center}
\includegraphics[scale=0.85]{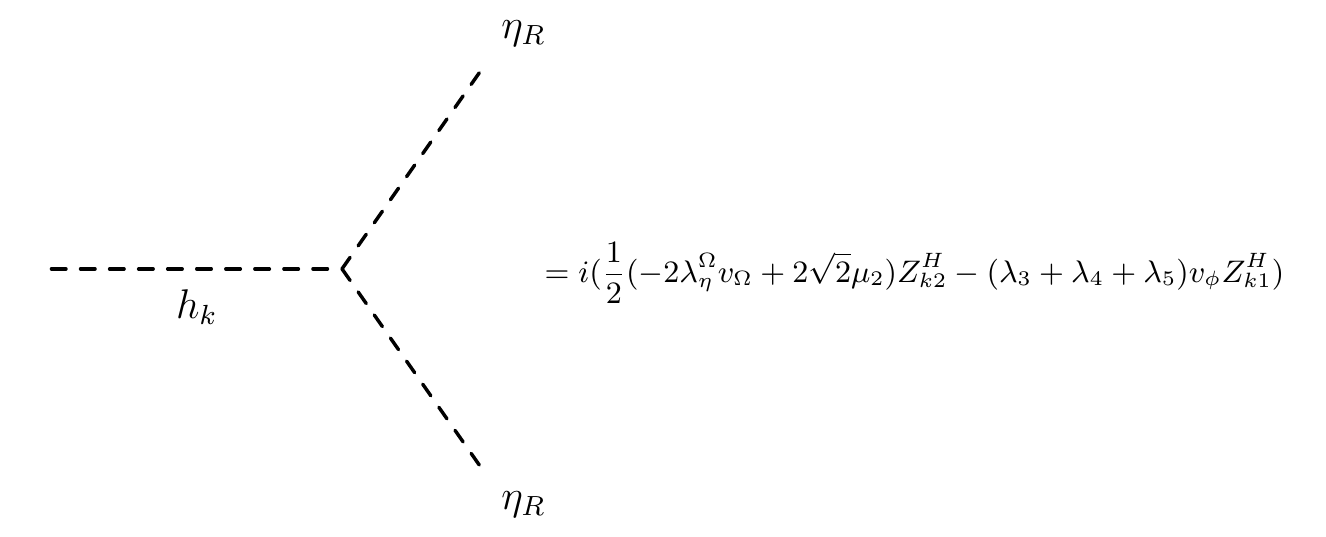}\vspace{0.5cm}
\includegraphics[scale=0.85]{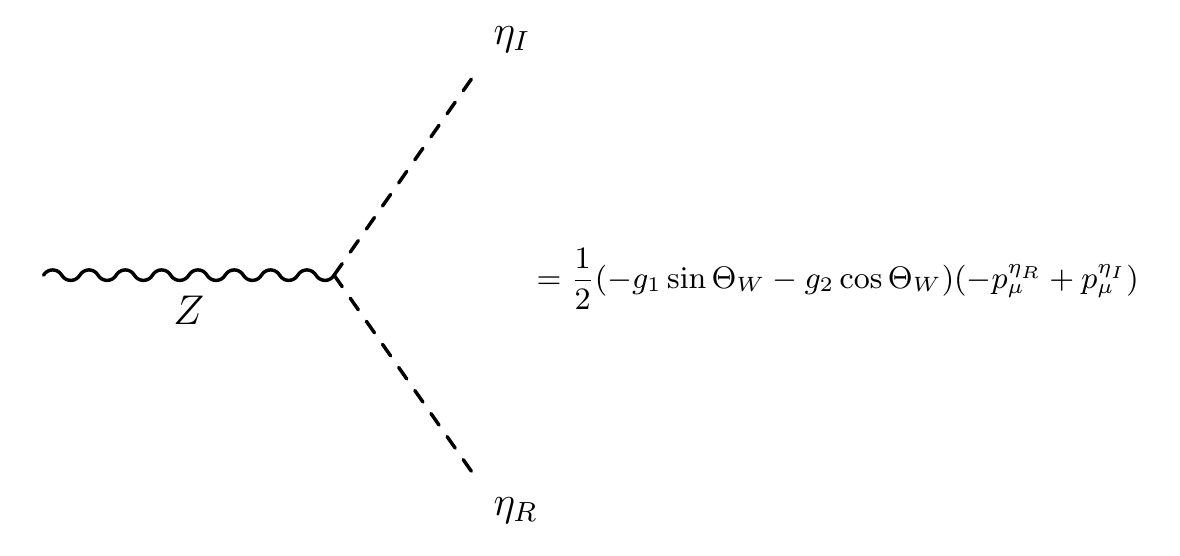}
\caption{Relevant $\eta_R$ interaction vertices. All particle momenta are considered as incoming. In the interaction with neutral scalars, $Z_{k1}^H$ and $Z_{k2}^H$ are entries of the mixing matrix that diagonalizes the mass matrix in Eq. \ref{neutral} ($k=1$ refers to the SM Higgs $h^0$ and $k=2$ to the heavy scalar $H$). In the interaction with the $Z^0$ boson, $g_1$ and $g_2$ are the electroweak coupling constants associated to the SM groups $U(1)_Y$ and $SU(2)_L$, respectively.}
\label{fig:feynman_rules}
\end{center}
\end{figure}

Figure~\ref{fig:feynman_rules} shows the most important Feynman rules for the relevant scalar dark-matter-physics interactions in the Singlet + Triplet Scotogenic Model.  
These are important for all the signatures studied in this paper, like the $\eta_R$-nucleon spin-independent elastic scattering and for the searches in the $\met+$jet channel at the LHC.

In contrast to the simplest Scotogenic Model, the interaction vertex with the Higgs is not fully determined by $\lambda_{345}$, as it contains an extra contribution dependent on 
$\lambda_{\eta}^{\Omega}$ and $\mu_2$, involving the heavy neutral scalar $H$, although weighted by its mixing with $h^0$.
Instead, the interaction vertex with the $Z^0$ boson depends on the quadrimomenta $p_\mu^{\eta_R}$, $p_\mu^{\eta_I}$ and on the electroweak couplings $g_1$ and $g_2$.

\acknowledgements 
\noindent

We are deeply grateful to Martin Hirsch and Avelino Vicente for valuable discussions and help with SARAH and SPheno. We also thank Nicolas Rojas and Sebastian Urrutia for their help with the Python numerical code. We further acknowledge useful discussions on ATLAS searches with Mar\'ia Moreno and Florencia Castillo. VDR is also grateful for the kind hospitality received at Fermilab during the final stage of this work. \\ Work supported by the Spanish grants FPA2017-85216-P, SEV-2014-0398 (MINECO/AEI/FEDER, UE), PROMETEO/2018/165 (Generalitat Valenciana) and FPA2017-90566-REDC (Red Consolider MultiDark).
IMA thanks the financial support by CONICYT, Doctorado Nacional (2015) N. 21151255. VDR acknowledges financial support by the  "Juan  de  la  Cierva  Incorporaci\'on" program  (IJCI-2016-27736) of the  Spanish MINECO and partial support by the EU Horizon 2020 project InvisiblesPlus (690575-InvisiblesPlus-H2020-MSCA-RISE-2015). LCD thanks "Coordenaç\~ao de Aperfeiçoamento de Pessoal de N\'ivel Superior - Brasil" (CAPES -Finance Code 001) for the financial support.

\bibliographystyle{utphys}

\begin{thebibliography}{10}

\bibitem{Schechter:1980gr}
J.~Schechter and J.~Valle, ``{Neutrino Masses in SU(2) x U(1) Theories},''
  \href{http://dx.doi.org/10.1103/PhysRevD.22.2227}{{\em Phys.Rev.} {\bfseries
  D22} (1980) 2227}.

\bibitem{Rojas:2018wym}
N.~Rojas, R.~Srivastava, and J.~W. Valle, ``{Simplest Scoto-Seesaw
  Mechanism},'' \href{http://dx.doi.org/10.1016/j.physletb.2018.12.014}{{\em
  Phys.Lett.} {\bfseries B789} (2019) 132--136},
  \href{http://arxiv.org/abs/1807.11447}{{\ttfamily arXiv:1807.11447
  [hep-ph]}}.

\bibitem{deSalas:2017kay}
P.~de~Salas, D.~Forero, C.~Ternes, M.~Tortola, and J.~Valle, ``{Status of
  neutrino oscillations 2018: 3$\sigma$ hint for normal mass ordering and
  improved CP sensitivity},''
  \href{http://dx.doi.org/10.1016/j.physletb.2018.06.019}{{\em Phys.Lett.}
  {\bfseries B782} (2018) 633--640},
  \href{http://arxiv.org/abs/1708.01186}{{\ttfamily arXiv:1708.01186
  [hep-ph]}}.

\bibitem{deSalas:2018bym}
P.~De~Salas, S.~Gariazzo, O.~Mena, C.~Ternes, and M.~T{\'o}rtola, ``{Neutrino
  Mass Ordering from Oscillations and Beyond: 2018 Status and Future
  Prospects},'' \href{http://dx.doi.org/10.3389/fspas.2018.00036}{{\em
  Front.Astron.Space Sci.} {\bfseries 5} (2018) 36},
  \href{http://arxiv.org/abs/1806.11051}{{\ttfamily arXiv:1806.11051
  [hep-ph]}}.

\bibitem{Dorame:2011eb}
L.~Dorame, D.~Meloni, S.~Morisi, E.~Peinado, and J.~Valle, ``{Constraining
  Neutrinoless Double Beta Decay},''
  \href{http://dx.doi.org/10.1016/j.nuclphysb.2012.04.003}{{\em Nucl.Phys.}
  {\bfseries B861} (2012) 259--270},
  \href{http://arxiv.org/abs/1111.5614}{{\ttfamily arXiv:1111.5614 [hep-ph]}}.

\bibitem{Dorame:2012zv}
L.~Dorame, S.~Morisi, E.~Peinado, J.~Valle, and A.~D. Rojas, ``{A new neutrino
  mass sum rule from inverse seesaw},''
  \href{http://dx.doi.org/10.1103/PhysRevD.86.056001}{{\em Phys.Rev.}
  {\bfseries D86} (2012) 056001},
  \href{http://arxiv.org/abs/1203.0155}{{\ttfamily arXiv:1203.0155 [hep-ph]}}.

\bibitem{King:2013hj}
S.~King, S.~Morisi, E.~Peinado, and J.~Valle, ``{Quark-Lepton Mass Relation in
  a Realistic $A_4$ Extension of the Standard Model},''
  \href{http://dx.doi.org/10.1016/j.physletb.2013.05.067}{{\em Phys.Lett.}
  {\bfseries B724} (2013) 68--72},
  \href{http://arxiv.org/abs/1301.7065}{{\ttfamily arXiv:1301.7065 [hep-ph]}}.

\bibitem{Ma:2006km}
E.~Ma, ``{Verifiable radiative seesaw mechanism of neutrino mass and dark
  matter},'' \href{http://dx.doi.org/10.1103/PhysRevD.73.077301}{{\em
  Phys.Rev.} {\bfseries D73} (2006) 077301}.

\bibitem{Hirsch:2013ola}
M.~Hirsch, R.~Lineros, S.~Morisi, J.~Palacio, N.~Rojas, and J.~Valle, ``{WIMP
  dark matter as radiative neutrino mass messenger},''
  \href{http://dx.doi.org/10.1007/JHEP10(2013)149}{{\em JHEP} {\bfseries 1310}
  (2013) 149}, \href{http://arxiv.org/abs/1307.8134}{{\ttfamily arXiv:1307.8134
  [hep-ph]}}.

\bibitem{Merle:2015ica}
A.~Merle and M.~Platscher, ``{Running of radiative neutrino masses: the
  scotogenic model {\textemdash} revisited},''
  \href{http://dx.doi.org/10.1007/JHEP11(2015)148}{{\em JHEP} {\bfseries 1511}
  (2015) 148}, \href{http://arxiv.org/abs/1507.06314}{{\ttfamily
  arXiv:1507.06314 [hep-ph]}}.

\bibitem{Merle:2016scw}
A.~Merle, M.~Platscher, N.~Rojas, J.~W.~F. Valle, and A.~Vicente,
  ``{Consistency of WIMP Dark Matter as radiative neutrino mass messenger},''
  \href{http://dx.doi.org/10.1007/JHEP07(2016)013}{{\em JHEP} {\bfseries 1607}
  (2016) 013}, \href{http://arxiv.org/abs/1603.05685}{{\ttfamily
  arXiv:1603.05685 [hep-ph]}}.

\bibitem{Diaz:2016udz}
M.~A. D{\'\i}az, N.~Rojas, S.~Urrutia-Quiroga, and J.~W.~F. Valle, ``{Heavy
  Higgs Boson Production at Colliders in the Singlet-Triplet Scotogenic Dark
  Matter Model},'' \href{http://dx.doi.org/10.1007/JHEP08(2017)017}{{\em JHEP}
  {\bfseries 1708} (2017) 017},
  \href{http://arxiv.org/abs/1612.06569}{{\ttfamily arXiv:1612.06569
  [hep-ph]}}.

\bibitem{Choubey:2017yyn}
S.~Choubey, S.~Khan, M.~Mitra, and S.~Mondal, ``{Singlet-Triplet Fermionic Dark
  Matter and LHC Phenomenology},''
  \href{http://dx.doi.org/10.1140/epjc/s10052-018-5785-1}{{\em Eur.Phys.J.}
  {\bfseries C78} (2018) 302},
  \href{http://arxiv.org/abs/1711.08888}{{\ttfamily arXiv:1711.08888
  [hep-ph]}}.

\bibitem{Restrepo:2019ilz}
D.~Restrepo and A.~Rivera, ``{Phenomenological consistency of the
  singlet-triplet scotogenic model},''
  \href{http://dx.doi.org/10.1007/JHEP04(2020)134}{{\em JHEP} {\bfseries 04}
  (2020) 134}, \href{http://arxiv.org/abs/1907.11938}{{\ttfamily
  arXiv:1907.11938 [hep-ph]}}.

\bibitem{Hagedorn:2018spx}
C.~Hagedorn, J.~Herrero-Garc{\'\i}a, E.~Molinaro, and M.~A. Schmidt,
  ``{Phenomenology of the Generalised Scotogenic Model with Fermionic Dark
  Matter},'' \href{http://dx.doi.org/10.1007/JHEP11(2018)103}{{\em JHEP}
  {\bfseries 1811} (2018) 103},
  \href{http://arxiv.org/abs/1804.04117}{{\ttfamily arXiv:1804.04117
  [hep-ph]}}.

\bibitem{Deshpande:1977rw}
N.~G. Deshpande and E.~Ma, ``{Pattern of Symmetry Breaking with Two Higgs
  Doublets},'' \href{http://dx.doi.org/10.1103/PhysRevD.18.2574}{{\em
  Phys.Rev.} {\bfseries D18} (1978) 2574}.

\bibitem{Barbieri:2006dq}
R.~Barbieri, L.~J. Hall, and V.~S. Rychkov, ``{Improved naturalness with a
  heavy Higgs: An Alternative road to LHC physics},''
  \href{http://dx.doi.org/10.1103/PhysRevD.74.015007}{{\em Phys.Rev.}
  {\bfseries D74} (2006) 015007}.

\bibitem{Diaz:2015pyv}
M.~A. D{\'\i}az, B.~Koch, and S.~Urrutia-Quiroga, ``{Constraints to Dark Matter
  from Inert Higgs Doublet Model},''
  \href{http://dx.doi.org/10.1155/2016/8278375}{{\em Adv.High Energy Phys.}
  {\bfseries 2016} (2016) 8278375},
  \href{http://arxiv.org/abs/1511.04429}{{\ttfamily arXiv:1511.04429
  [hep-ph]}}.

\bibitem{Honorez:2010re}
L.~Lopez~Honorez and C.~E. Yaguna, ``{The inert doublet model of dark matter
  revisited},'' \href{http://dx.doi.org/10.1007/JHEP09(2010)046}{{\em JHEP}
  {\bfseries 1009} (2010) 046},
  \href{http://arxiv.org/abs/1003.3125}{{\ttfamily arXiv:1003.3125 [hep-ph]}}.

\bibitem{LopezHonorez:2006gr}
L.~Lopez~Honorez, E.~Nezri, J.~F. Oliver, and M.~H. Tytgat, ``{The Inert
  Doublet Model: An Archetype for Dark Matter},''
  \href{http://dx.doi.org/10.1088/1475-7516/2007/02/028}{{\em JCAP} {\bfseries
  0702} (2007) 028}.

\bibitem{Hirsch:2010ru}
M.~Hirsch, S.~Morisi, E.~Peinado, and J.~Valle, ``{Discrete dark matter},''
  \href{http://dx.doi.org/10.1103/PhysRevD.82.116003}{{\em Phys.Rev.}
  {\bfseries D82} (2010) 116003},
  \href{http://arxiv.org/abs/1007.0871}{{\ttfamily arXiv:1007.0871 [hep-ph]}}.

\bibitem{Boucenna:2012qb}
M.~Boucenna, S.~Morisi, E.~Peinado, Y.~Shimizu, and J.~Valle, ``{Predictive
  discrete dark matter model and neutrino oscillations},''
  \href{http://dx.doi.org/10.1103/PhysRevD.86.073008}{{\em Phys.Rev.}
  {\bfseries D86} (2012) 073008},
  \href{http://arxiv.org/abs/1204.4733}{{\ttfamily arXiv:1204.4733 [hep-ph]}}.

\bibitem{Gunion:1989ci}
J.~Gunion, R.~Vega, and J.~Wudka, ``{Higgs triplets in the standard model},''
  \href{http://dx.doi.org/10.1103/PhysRevD.42.1673}{{\em Phys.Rev.} {\bfseries
  D42} (1990) 1673--1691}.

\bibitem{Gunion:1989we}
J.~F. Gunion, H.~E. Haber, G.~L. Kane, and S.~Dawson, ``{The Higgs Hunter's
  Guide},'' {\em Front.Phys.} {\bfseries 80} (2000) 1--404.

\bibitem{tHooft:1979rat}
G.~'t~Hooft, ``{Naturalness, chiral symmetry, and spontaneous chiral symmetry
  breaking},'' \href{http://dx.doi.org/10.1007/978-1-4684-7571-5_9}{{\em NATO
  Sci. Ser. B} {\bfseries 59} (1980) 135--157}.

\bibitem{Passarino:1978jh}
G.~Passarino and M.~Veltman, ``{One Loop Corrections for e+ e- Annihilation
  Into mu+ mu- in the Weinberg Model},''
  \href{http://dx.doi.org/10.1016/0550-3213(79)90234-7}{{\em Nucl.Phys.}
  {\bfseries B160} (1979) 151--207}.

\bibitem{Casas:2001sr}
J.~Casas and A.~Ibarra, ``{Oscillating neutrinos and muon ---> e, gamma},''
  \href{http://dx.doi.org/10.1016/S0550-3213(01)00475-8}{{\em Nucl.Phys.}
  {\bfseries B618} (2001) 171--204}.

\bibitem{Rodejohann:2011vc}
W.~Rodejohann and J.~Valle, ``{Symmetrical Parametrizations of the Lepton
  Mixing Matrix},'' \href{http://dx.doi.org/10.1103/PhysRevD.84.073011}{{\em
  Phys.Rev.} {\bfseries D84} (2011) 073011},
  \href{http://arxiv.org/abs/1108.3484}{{\ttfamily arXiv:1108.3484 [hep-ph]}}.

\bibitem{Reig:2018ztc}
M.~Reig, D.~Restrepo, J.~F. Valle, and O.~Zapata, ``{Bound-state dark matter
  with Majorana neutrinos},''
  \href{http://dx.doi.org/10.1016/j.physletb.2019.01.023}{{\em Phys.Lett.}
  {\bfseries B790} (2019) 303--307},
  \href{http://arxiv.org/abs/1806.09977}{{\ttfamily arXiv:1806.09977
  [hep-ph]}}.

\bibitem{Leite:2019grf}
J.~Leite, O.~Popov, R.~Srivastava, and J.~W. Valle, ``{A theory for scotogenic
  dark matter stabilised by residual gauge symmetry},''
  \href{http://arxiv.org/abs/1909.06386}{{\ttfamily arXiv:1909.06386
  [hep-ph]}}.

\bibitem{Alduino:2017ehq}
{\bfseries CUORE} Collaboration, C.~Alduino {\em et~al.}, ``{First Results from
  CUORE: A Search for Lepton Number Violation via $0\nu\beta\beta$ Decay of
  $^{130}$Te},'' \href{http://dx.doi.org/10.1103/PhysRevLett.120.132501}{{\em
  Phys.Rev.Lett.} {\bfseries 120} (2018) 132501},
  \href{http://arxiv.org/abs/1710.07988}{{\ttfamily arXiv:1710.07988
  [nucl-ex]}}.

\bibitem{Albert:2017owj}
{\bfseries EXO} Collaboration, J.~Albert {\em et~al.}, ``{Search for
  Neutrinoless Double-Beta Decay with the Upgraded EXO-200 Detector},''
  \href{http://dx.doi.org/10.1103/PhysRevLett.120.072701}{{\em Phys.Rev.Lett.}
  {\bfseries 120} (2018) 072701},
  \href{http://arxiv.org/abs/1707.08707}{{\ttfamily arXiv:1707.08707
  [hep-ex]}}.

\bibitem{Agostini:2018tnm}
{\bfseries GERDA} Collaboration, M.~Agostini {\em et~al.}, ``{Improved Limit on
  Neutrinoless Double-$\beta$ Decay of $^{76}$Ge from GERDA Phase II},''
  \href{http://dx.doi.org/10.1103/PhysRevLett.120.132503}{{\em Phys.Rev.Lett.}
  {\bfseries 120} (2018) 132503},
  \href{http://arxiv.org/abs/1803.11100}{{\ttfamily arXiv:1803.11100
  [nucl-ex]}}.

\bibitem{KamLAND-Zen:2016pfg}
{\bfseries KamLAND-Zen} Collaboration, A.~Gando {\em et~al.}, ``{Search for
  Majorana Neutrinos near the Inverted Mass Hierarchy Region with
  KamLAND-Zen},'' \href{http://dx.doi.org/10.1103/PhysRevLett.117.109903}{{\em
  Phys.Rev.Lett.} {\bfseries 117} (2016) 082503},
  \href{http://arxiv.org/abs/1605.02889}{{\ttfamily arXiv:1605.02889
  [hep-ex]}}.

\bibitem{Andringa:2015tza}
{\bfseries SNO+} Collaboration, S.~Andringa {\em et~al.}, ``{Current Status and
  Future Prospects of the SNO+ Experiment},''
  \href{http://dx.doi.org/10.1155/2016/6194250}{{\em Adv.High Energy Phys.}
  {\bfseries 2016} (2016) 6194250},
  \href{http://arxiv.org/abs/1508.05759}{{\ttfamily arXiv:1508.05759
  [physics.ins-det]}}.

\bibitem{Abgrall:2017syy}
{\bfseries LEGEND} Collaboration, N.~Abgrall {\em et~al.},
  \href{http://dx.doi.org/10.1063/1.5007652}{``{The Large Enriched Germanium
  Experiment for Neutrinoless Double Beta Decay (LEGEND)},''} vol.~1894,
  p.~020027.
\newblock 2017.
\newblock \href{http://arxiv.org/abs/1709.01980}{{\ttfamily arXiv:1709.01980
  [physics.ins-det]}}.

\bibitem{Albert:2017hjq}
{\bfseries nEXO} Collaboration, J.~Albert {\em et~al.}, ``{Sensitivity and
  Discovery Potential of nEXO to Neutrinoless Double Beta Decay},''
  \href{http://dx.doi.org/10.1103/PhysRevC.97.065503}{{\em Phys.Rev.}
  {\bfseries C97} (2018) 065503},
  \href{http://arxiv.org/abs/1710.05075}{{\ttfamily arXiv:1710.05075
  [nucl-ex]}}.

\bibitem{Staub:2013tta}
F.~Staub, ``{SARAH 4 : A tool for (not only SUSY) model builders},''
  \href{http://dx.doi.org/10.1016/j.cpc.2014.02.018}{{\em Comput.Phys.Commun.}
  {\bfseries 185} (2014) 1773--1790},
  \href{http://arxiv.org/abs/1309.7223}{{\ttfamily arXiv:1309.7223 [hep-ph]}}.

\bibitem{Staub:2015kfa}
F.~Staub, ``{Exploring new models in all detail with SARAH},''
  \href{http://dx.doi.org/10.1155/2015/840780}{{\em Adv.High Energy Phys.}
  {\bfseries 2015} (2015) 840780},
  \href{http://arxiv.org/abs/1503.04200}{{\ttfamily arXiv:1503.04200
  [hep-ph]}}.

\bibitem{Porod:2003um}
W.~Porod, ``{SPheno, a program for calculating supersymmetric spectra, SUSY
  particle decays and SUSY particle production at e+ e- colliders},''
  \href{http://dx.doi.org/10.1016/S0010-4655(03)00222-4}{{\em
  Comput.Phys.Commun.} {\bfseries 153} (2003) 275--315}.

\bibitem{Porod:2011nf}
W.~Porod and F.~Staub, ``{SPheno 3.1: Extensions including flavour, CP-phases
  and models beyond the MSSM},''
  \href{http://dx.doi.org/10.1016/j.cpc.2012.05.021}{{\em Comput.Phys.Commun.}
  {\bfseries 183} (2012) 2458--2469},
  \href{http://arxiv.org/abs/1104.1573}{{\ttfamily arXiv:1104.1573 [hep-ph]}}.

\bibitem{Porod:2014xia}
W.~Porod, F.~Staub, and A.~Vicente, ``{A Flavor Kit for BSM models},''
  \href{http://dx.doi.org/10.1140/epjc/s10052-014-2992-2}{{\em Eur.Phys.J.}
  {\bfseries C74} (2014) 2992},
  \href{http://arxiv.org/abs/1405.1434}{{\ttfamily arXiv:1405.1434 [hep-ph]}}.

\bibitem{Belanger:2014vza}
G.~B{\'e}langer, F.~Boudjema, A.~Pukhov, and A.~Semenov, ``{micrOMEGAs4.1: two
  dark matter candidates},''
  \href{http://dx.doi.org/10.1016/j.cpc.2015.03.003}{{\em Comput.Phys.Commun.}
  {\bfseries 192} (2015) 322--329},
  \href{http://arxiv.org/abs/1407.6129}{{\ttfamily arXiv:1407.6129 [hep-ph]}}.

\bibitem{Alwall:2014hca}
J.~Alwall {\em et~al.}, ``{The automated computation of tree-level and
  next-to-leading order differential cross sections, and their matching to
  parton shower simulations},''
  \href{http://dx.doi.org/10.1007/JHEP07(2014)079}{{\em JHEP} {\bfseries 1407}
  (2014) 079}, \href{http://arxiv.org/abs/1405.0301}{{\ttfamily arXiv:1405.0301
  [hep-ph]}}.

\bibitem{PhysRevD.9.3320}
L.~Dolan and R.~Jackiw, ``Symmetry behavior at finite temperature,''
  \href{http://dx.doi.org/10.1103/PhysRevD.9.3320}{{\em Phys. Rev. D}
  {\bfseries 9} (Jun, 1974) 3320--3341}.
  \url{https://link.aps.org/doi/10.1103/PhysRevD.9.3320}.

\bibitem{PhysRevD.45.2933}
M.~E. Carrington, ``Effective potential at finite temperature in the standard
  model,'' \href{http://dx.doi.org/10.1103/PhysRevD.45.2933}{{\em Phys. Rev. D}
  {\bfseries 45} (Apr, 1992) 2933--2944}.
  \url{https://link.aps.org/doi/10.1103/PhysRevD.45.2933}.

\bibitem{Rocha-Moran:2016enp}
P.~Rocha-Moran and A.~Vicente, ``{Lepton Flavor Violation in the
  singlet-triplet scotogenic model},''
  \href{http://dx.doi.org/10.1007/JHEP07(2016)078}{{\em JHEP} {\bfseries 1607}
  (2016) 078}, \href{http://arxiv.org/abs/1605.01915}{{\ttfamily
  arXiv:1605.01915 [hep-ph]}}.

\bibitem{TheMEG:2016wtm}
{\bfseries MEG} Collaboration, A.~Baldini {\em et~al.}, ``{Search for the
  lepton flavour violating decay $\mu ^+ \rightarrow \mathrm {e}^+ \gamma $
  with the full dataset of the MEG experiment},''
  \href{http://dx.doi.org/10.1140/epjc/s10052-016-4271-x}{{\em Eur.Phys.J.}
  {\bfseries C76} (2016) 434},
  \href{http://arxiv.org/abs/1605.05081}{{\ttfamily arXiv:1605.05081
  [hep-ex]}}.

\bibitem{Bellgardt:1987du}
{\bfseries SINDRUM} Collaboration, U.~Bellgardt {\em et~al.}, ``{Search for the
  Decay mu+ ---> e+ e+ e-},''
  \href{http://dx.doi.org/10.1016/0550-3213(88)90462-2}{{\em Nucl.Phys.}
  {\bfseries B299} (1988) 1--6}.

\bibitem{Bertl:2006up}
{\bfseries SINDRUM II} Collaboration, W.~H. Bertl {\em et~al.}, ``{A Search for
  muon to electron conversion in muonic gold},''
  \href{http://dx.doi.org/10.1140/epjc/s2006-02582-x}{{\em Eur.Phys.J.}
  {\bfseries C47} (2006) 337--346}.

\bibitem{Peskin:1991sw}
M.~E. Peskin and T.~Takeuchi, ``{Estimation of oblique electroweak
  corrections},'' \href{http://dx.doi.org/10.1103/PhysRevD.46.381}{{\em
  Phys.Rev.} {\bfseries D46} (1992) 381--409}.

\bibitem{Abada:2018zra}
A.~Abada and T.~Toma, ``{Electric Dipole Moments in the Minimal Scotogenic
  Model},'' \href{http://dx.doi.org/10.1007/JHEP04(2018)030}{{\em JHEP}
  {\bfseries 1804} (2018) 030},
  \href{http://arxiv.org/abs/1802.00007}{{\ttfamily arXiv:1802.00007
  [hep-ph]}}.

\bibitem{Tanabashi:2018oca}
{\bfseries Particle Data Group} Collaboration, M.~Tanabashi {\em et~al.},
  ``{Review of Particle Physics},''
  \href{http://dx.doi.org/10.1103/PhysRevD.98.030001}{{\em Phys.Rev.}
  {\bfseries D98} (2018) 030001}.

\bibitem{Ade:2015xua}
{\bfseries Planck} Collaboration, P.~Ade {\em et~al.}, ``{Planck 2015 results.
  XIII. Cosmological parameters},''
  \href{http://dx.doi.org/10.1051/0004-6361/201525830}{{\em Astron.Astrophys.}
  {\bfseries 594} (2016) A13},
  \href{http://arxiv.org/abs/1502.01589}{{\ttfamily arXiv:1502.01589
  [astro-ph.CO]}}.

\bibitem{Aghanim:2018eyx}
{\bfseries Planck} Collaboration, N.~Aghanim {\em et~al.}, ``{Planck 2018
  results. VI. Cosmological parameters},''
  \href{http://arxiv.org/abs/1807.06209}{{\ttfamily arXiv:1807.06209
  [astro-ph.CO]}}.

\bibitem{Aprile:2018dbl}
{\bfseries XENON} Collaboration, E.~Aprile {\em et~al.}, ``{Dark Matter Search
  Results from a One Ton-Year Exposure of XENON1T},''
  \href{http://dx.doi.org/10.1103/PhysRevLett.121.111302}{{\em Phys.Rev.Lett.}
  {\bfseries 121} (2018) 111302},
  \href{http://arxiv.org/abs/1805.12562}{{\ttfamily arXiv:1805.12562
  [astro-ph.CO]}}.

\bibitem{Hambye:2009pw}
T.~Hambye, F.-S. Ling, L.~Lopez~Honorez, and J.~Rocher, ``{Scalar Multiplet
  Dark Matter},'' \href{http://dx.doi.org/10.1088/1126-6708/2009/07/090}{{\em
  JHEP} {\bfseries 0907} (2009) 090},
  \href{http://arxiv.org/abs/0903.4010}{{\ttfamily arXiv:0903.4010 [hep-ph]}}.

\bibitem{Akerib:2016vxi}
{\bfseries LUX} Collaboration, D.~Akerib {\em et~al.}, ``{Results from a search
  for dark matter in the complete LUX exposure},''
  \href{http://dx.doi.org/10.1103/PhysRevLett.118.021303}{{\em Phys.Rev.Lett.}
  {\bfseries 118} (2017) 021303},
  \href{http://arxiv.org/abs/1608.07648}{{\ttfamily arXiv:1608.07648
  [astro-ph.CO]}}.

\bibitem{Cui:2017nnn}
{\bfseries PandaX-II} Collaboration, X.~Cui {\em et~al.}, ``{Dark Matter
  Results From 54-Ton-Day Exposure of PandaX-II Experiment},''
  \href{http://dx.doi.org/10.1103/PhysRevLett.119.181302}{{\em Phys.Rev.Lett.}
  {\bfseries 119} (2017) 181302},
  \href{http://arxiv.org/abs/1708.06917}{{\ttfamily arXiv:1708.06917
  [astro-ph.CO]}}.

\bibitem{Agnes:2018ves}
{\bfseries DarkSide} Collaboration, P.~Agnes {\em et~al.}, ``{Low-Mass Dark
  Matter Search with the DarkSide-50 Experiment},''
  \href{http://dx.doi.org/10.1103/PhysRevLett.121.081307}{{\em Phys.Rev.Lett.}
  {\bfseries 121} (2018) 081307},
  \href{http://arxiv.org/abs/1802.06994}{{\ttfamily arXiv:1802.06994
  [astro-ph.HE]}}.

\bibitem{Ajaj:2019imk}
{\bfseries DEAP} Collaboration, R.~Ajaj {\em et~al.}, ``{Search for dark matter
  with a 231-day exposure of liquid argon using DEAP-3600 at SNOLAB},''
  \href{http://dx.doi.org/10.1103/PhysRevD.100.022004}{{\em Phys.Rev.}
  {\bfseries D100} (2019) 022004},
  \href{http://arxiv.org/abs/1902.04048}{{\ttfamily arXiv:1902.04048
  [astro-ph.CO]}}.

\bibitem{Billard:2013qya}
J.~Billard, L.~Strigari, and E.~Figueroa-Feliciano, ``{Implication of neutrino
  backgrounds on the reach of next generation dark matter direct detection
  experiments},'' \href{http://dx.doi.org/10.1103/PhysRevD.89.023524}{{\em
  Phys.Rev.} {\bfseries D89} (2014) 023524},
  \href{http://arxiv.org/abs/1307.5458}{{\ttfamily arXiv:1307.5458 [hep-ph]}}.

\bibitem{Akerib:2018lyp}
{\bfseries LUX-ZEPLIN} Collaboration, D.~S. Akerib {\em et~al.}, ``{Projected
  WIMP Sensitivity of the LUX-ZEPLIN (LZ) Dark Matter Experiment},''
  \href{http://arxiv.org/abs/1802.06039}{{\ttfamily arXiv:1802.06039
  [astro-ph.IM]}}.

\bibitem{Ackermann:2015zua}
{\bfseries Fermi-LAT} Collaboration, M.~Ackermann {\em et~al.}, ``{Searching
  for Dark Matter Annihilation from Milky~Way Dwarf Spheroidal Galaxies with
  Six Years of Fermi Large Area Telescope Data},''
  \href{http://dx.doi.org/10.1103/PhysRevLett.115.231301}{{\em Phys.Rev.Lett.}
  {\bfseries 115} (2015) 231301},
  \href{http://arxiv.org/abs/1503.02641}{{\ttfamily arXiv:1503.02641
  [astro-ph.HE]}}.

\bibitem{Abdallah:2016ygi}
{\bfseries H.E.S.S.} Collaboration, H.~Abdallah {\em et~al.}, ``{Search for
  dark matter annihilations towards the inner Galactic halo from 10 years of
  observations with H.E.S.S},''
  \href{http://dx.doi.org/10.1103/PhysRevLett.117.111301}{{\em Phys.Rev.Lett.}
  {\bfseries 117} (2016) 111301},
  \href{http://arxiv.org/abs/1607.08142}{{\ttfamily arXiv:1607.08142
  [astro-ph.HE]}}.

\bibitem{Charles:2016pgz}
{\bfseries Fermi-LAT} Collaboration, E.~Charles {\em et~al.}, ``{Sensitivity
  Projections for Dark Matter Searches with the Fermi Large Area Telescope},''
  \href{http://dx.doi.org/10.1016/j.physrep.2016.05.001}{{\em Phys.Rept.}
  {\bfseries 636} (2016) 1--46},
  \href{http://arxiv.org/abs/1605.02016}{{\ttfamily arXiv:1605.02016
  [astro-ph.HE]}}.

\bibitem{Acharya:2017ttl}
{\bfseries CTA Consortium} Collaboration, B.~Acharya {\em et~al.},
  \href{http://dx.doi.org/10.1142/10986}{{\em {Science with the Cherenkov
  Telescope Array}}}.
\newblock 2017.
\newblock \href{http://arxiv.org/abs/1709.07997}{{\ttfamily arXiv:1709.07997
  [astro-ph.IM]}}.

\bibitem{1931AnP...403..257S}
A.~{Sommerfeld}, ``{{\"U}ber die Beugung und Bremsung der Elektronen},''
  \href{http://dx.doi.org/10.1002/andp.19314030302}{{\em Annalen der Physik}
  {\bfseries 403} (1931) 257--330}.


\bibitem{Hisano:2003ec}
J.~Hisano, S.~Matsumoto, and M.~M. Nojiri, ``{Explosive dark matter
  annihilation},'' \href{http://dx.doi.org/10.1103/PhysRevLett.92.031303}{{\em
  Phys.Rev.Lett.} {\bfseries 92} (2004) 031303}.

\bibitem{Hisano:2004ds}
J.~Hisano, S.~Matsumoto, M.~M. Nojiri, and O.~Saito, ``{Non-perturbative effect
  on dark matter annihilation and gamma ray signature from galactic center},''
  \href{http://dx.doi.org/10.1103/PhysRevD.71.063528}{{\em Phys.Rev.}
  {\bfseries D71} (2005) 063528}.

\bibitem{ArkaniHamed:2008qn}
N.~Arkani-Hamed, D.~P. Finkbeiner, T.~R. Slatyer, and N.~Weiner, ``{A Theory of
  Dark Matter},'' \href{http://dx.doi.org/10.1103/PhysRevD.79.015014}{{\em
  Phys.Rev.} {\bfseries D79} (2009) 015014},
  \href{http://arxiv.org/abs/0810.0713}{{\ttfamily arXiv:0810.0713 [hep-ph]}}.

\bibitem{Chowdhury:2016mtl}
T.~A. Chowdhury and S.~Nasri, ``{The Sommerfeld Enhancement in the Scotogenic
  Model with Large Electroweak Scalar Multiplets},''
  \href{http://dx.doi.org/10.1088/1475-7516/2017/01/041}{{\em JCAP} {\bfseries
  1701} (2017) 041}, \href{http://arxiv.org/abs/1611.06590}{{\ttfamily
  arXiv:1611.06590 [hep-ph]}}.

\bibitem{Adriani:2008zr}
{\bfseries PAMELA} Collaboration, O.~Adriani {\em et~al.}, ``{An anomalous
  positron abundance in cosmic rays with energies 1.5-100 GeV},''
  \href{http://dx.doi.org/10.1038/nature07942}{{\em Nature} {\bfseries 458}
  (2009) 607--609}, \href{http://arxiv.org/abs/0810.4995}{{\ttfamily
  arXiv:0810.4995 [astro-ph]}}.

\bibitem{Adriani:2013uda}
{\bfseries PAMELA} Collaboration, O.~Adriani {\em et~al.}, ``{Cosmic-Ray
  Positron Energy Spectrum Measured by PAMELA},''
  \href{http://dx.doi.org/10.1103/PhysRevLett.111.081102}{{\em Phys.Rev.Lett.}
  {\bfseries 111} (2013) 081102},
  \href{http://arxiv.org/abs/1308.0133}{{\ttfamily arXiv:1308.0133
  [astro-ph.HE]}}.

\bibitem{Aguilar:2013qda}
{\bfseries AMS} Collaboration, M.~Aguilar {\em et~al.}, ``{First Result from
  the Alpha Magnetic Spectrometer on the International Space Station: Precision
  Measurement of the Positron Fraction in Primary Cosmic Rays of
  0.5{\textendash}350 GeV},''
  \href{http://dx.doi.org/10.1103/PhysRevLett.110.141102}{{\em Phys.Rev.Lett.}
  {\bfseries 110} (2013) 141102}.

\bibitem{Bergstrom:2013jra}
L.~Bergstrom, T.~Bringmann, I.~Cholis, D.~Hooper, and C.~Weniger, ``{New Limits
  on Dark Matter Annihilation from AMS Cosmic Ray Positron Data},''
  \href{http://dx.doi.org/10.1103/PhysRevLett.111.171101}{{\em Phys.Rev.Lett.}
  {\bfseries 111} (2013) 171101},
  \href{http://arxiv.org/abs/1306.3983}{{\ttfamily arXiv:1306.3983
  [astro-ph.HE]}}.

\bibitem{Aguilar:2016kjl}
{\bfseries AMS} Collaboration, M.~Aguilar {\em et~al.}, ``{Antiproton Flux,
  Antiproton-to-Proton Flux Ratio, and Properties of Elementary Particle Fluxes
  in Primary Cosmic Rays Measured with the Alpha Magnetic Spectrometer on the
  International Space Station},''
  \href{http://dx.doi.org/10.1103/PhysRevLett.117.091103}{{\em Phys.Rev.Lett.}
  {\bfseries 117} (2016) 091103}.

\bibitem{Cuoco:2016eej}
A.~Cuoco, M.~Kr{\"a}mer, and M.~Korsmeier, ``{Novel Dark Matter Constraints
  from Antiprotons in Light of AMS-02},''
  \href{http://dx.doi.org/10.1103/PhysRevLett.118.191102}{{\em Phys.Rev.Lett.}
  {\bfseries 118} (2017) 191102},
  \href{http://arxiv.org/abs/1610.03071}{{\ttfamily arXiv:1610.03071
  [astro-ph.HE]}}.

\bibitem{Cholis:2019ejx}
I.~Cholis, T.~Linden, and D.~Hooper, ``{A Robust Excess in the Cosmic-Ray
  Antiproton Spectrum: Implications for Annihilating Dark Matter},''
  \href{http://dx.doi.org/10.1103/PhysRevD.99.103026}{{\em Phys.Rev.}
  {\bfseries D99} (2019) 103026},
  \href{http://arxiv.org/abs/1903.02549}{{\ttfamily arXiv:1903.02549
  [astro-ph.HE]}}.

\bibitem{Donato:1999gy}
F.~Donato, N.~Fornengo, and P.~Salati, ``{Anti-deuterons as a signature of
  supersymmetric dark matter},''
  \href{http://dx.doi.org/10.1103/PhysRevD.62.043003}{{\em Phys.Rev.}
  {\bfseries D62} (2000) 043003}.

\bibitem{Carlson:2014ssa}
E.~Carlson, A.~Coogan, T.~Linden, S.~Profumo, A.~Ibarra, and S.~Wild,
  ``{Antihelium from Dark Matter},''
  \href{http://dx.doi.org/10.1103/PhysRevD.89.076005}{{\em Phys.Rev.}
  {\bfseries D89} (2014) 076005},
  \href{http://arxiv.org/abs/1401.2461}{{\ttfamily arXiv:1401.2461 [hep-ph]}}.

\bibitem{Cirelli:2014qia}
M.~Cirelli, N.~Fornengo, M.~Taoso, and A.~Vittino, ``{Anti-helium from Dark
  Matter annihilations},''
  \href{http://dx.doi.org/10.1007/JHEP08(2014)009}{{\em JHEP} {\bfseries 1408}
  (2014) 009}, \href{http://arxiv.org/abs/1401.4017}{{\ttfamily arXiv:1401.4017
  [hep-ph]}}.

\bibitem{Coogan:2017pwt}
A.~Coogan and S.~Profumo, ``{Origin of the tentative AMS antihelium events},''
  \href{http://dx.doi.org/10.1103/PhysRevD.96.083020}{{\em Phys. Rev.}
  {\bfseries D96} no.~8, (2017) 083020},
  \href{http://arxiv.org/abs/1705.09664}{{\ttfamily arXiv:1705.09664
  [astro-ph.HE]}}.

\bibitem{ATLAS:2018ghb}
{\bfseries ATLAS} Collaboration, ``{$E_{\text{T}}^{\text{miss}}$ performance in
  the ATLAS detector using 2015-2016 LHC p-p collisions},''.

\bibitem{CMS:2016ljj}
{\bfseries CMS} Collaboration, ``{Performance of missing energy reconstruction
  in 13 TeV pp collision data using the CMS detector},''.

\bibitem{Aaboud:2016tnv}
{\bfseries ATLAS} Collaboration, M.~Aaboud {\em et~al.}, ``{Search for new
  phenomena in final states with an energetic jet and large missing transverse
  momentum in $pp$ collisions at $\sqrt{s}=13$ TeV using the ATLAS detector},''
  \href{http://dx.doi.org/10.1103/PhysRevD.94.032005}{{\em Phys.Rev.}
  {\bfseries D94} (2016) 032005},
  \href{http://arxiv.org/abs/1604.07773}{{\ttfamily arXiv:1604.07773
  [hep-ex]}}.

\bibitem{Haisch:2018hbm}
U.~Haisch and G.~Polesello, ``{Searching for dark matter in final states with
  two jets and missing transverse energy},''
  \href{http://dx.doi.org/10.1007/JHEP02(2019)128}{{\em JHEP} {\bfseries 1902}
  (2019) 128}, \href{http://arxiv.org/abs/1812.08129}{{\ttfamily
  arXiv:1812.08129 [hep-ph]}}.

\bibitem{Aaboud:2018ujj}
{\bfseries ATLAS} Collaboration, M.~Aaboud {\em et~al.}, ``{Search for new
  phenomena using the invariant mass distribution of same-flavour opposite-sign
  dilepton pairs in events with missing transverse momentum in $\sqrt{s}=13$ ~
  $\text {Te}\text {V}$ pp collisions with the ATLAS detector},''
  \href{http://dx.doi.org/10.1140/epjc/s10052-018-6081-9}{{\em Eur.Phys.J.}
  {\bfseries C78} (2018) 625},
  \href{http://arxiv.org/abs/1805.11381}{{\ttfamily arXiv:1805.11381
  [hep-ex]}}.

\bibitem{Aaboud:2017dor}
{\bfseries ATLAS} Collaboration, M.~Aaboud {\em et~al.}, ``{Search for dark
  matter at $\sqrt{s}=13$ TeV in final states containing an energetic photon
  and large missing transverse momentum with the ATLAS detector},''
  \href{http://dx.doi.org/10.1140/epjc/s10052-017-4965-8}{{\em Eur.Phys.J.}
  {\bfseries C77} (2017) 393},
  \href{http://arxiv.org/abs/1704.03848}{{\ttfamily arXiv:1704.03848
  [hep-ex]}}.

\bibitem{CMS}
``{Measurements of the Higgs boson production and decay rates and constraints
  on its couplings from a combined ATLAS and CMS analysis of the LHC pp
  collision data at $\sqrt{s}$ = 7 and 8 TeV},''.

\bibitem{Dercks:2016npn}
D.~Dercks, N.~Desai, J.~S. Kim, K.~Rolbiecki, J.~Tattersall, and T.~Weber,
  ``{CheckMATE 2: From the model to the limit},''
  \href{http://dx.doi.org/10.1016/j.cpc.2017.08.021}{{\em Comput.Phys.Commun.}
  {\bfseries 221} (2017) 383--418},
  \href{http://arxiv.org/abs/1611.09856}{{\ttfamily arXiv:1611.09856
  [hep-ph]}}.

\bibitem{Sjostrand:2007gs}
T.~Sjostrand, S.~Mrenna, and P.~Z. Skands, ``{A Brief Introduction to PYTHIA
  8.1},'' \href{http://dx.doi.org/10.1016/j.cpc.2008.01.036}{{\em
  Comput.Phys.Commun.} {\bfseries 178} (2008) 852--867},
  \href{http://arxiv.org/abs/0710.3820}{{\ttfamily arXiv:0710.3820 [hep-ph]}}.

\bibitem{deFavereau:2013fsa}
{\bfseries DELPHES 3} Collaboration, J.~de~Favereau {\em et~al.}, ``{DELPHES 3,
  A modular framework for fast simulation of a generic collider experiment},''
  \href{http://dx.doi.org/10.1007/JHEP02(2014)057}{{\em JHEP} {\bfseries 1402}
  (2014) 057}, \href{http://arxiv.org/abs/1307.6346}{{\ttfamily arXiv:1307.6346
  [hep-ex]}}.

\bibitem{Cacciari:2011ma}
M.~Cacciari, G.~P. Salam, and G.~Soyez, ``{FastJet User Manual},''
  \href{http://dx.doi.org/10.1140/epjc/s10052-012-1896-2}{{\em Eur.Phys.J.}
  {\bfseries C72} (2012) 1896},
  \href{http://arxiv.org/abs/1111.6097}{{\ttfamily arXiv:1111.6097 [hep-ph]}}.

\bibitem{Read:2002hq}
A.~L. Read,
  \href{http://dx.doi.org/10.1088/0954-3899/28/10/313}{``{Presentation of
  search results: The CL(s) technique},''} vol.~G28, pp.~2693--2704.
\newblock 2002.

\bibitem{Domingo:2018ykx}
F.~Domingo, J.~S. Kim, V.~Martin-Lozano, P.~Martin-Ramiro, and R.~Ruiz~de
  Austri, ``{Confronting the neutralino and chargino sector of the NMSSM to the
  multi-lepton searches at the LHC},''
  \href{http://arxiv.org/abs/1812.05186}{{\ttfamily arXiv:1812.05186
  [hep-ph]}}.

\bibitem{Belyaev:2018ext}
A.~Belyaev {\em et~al.}, ``{Advancing LHC probes of dark matter from the inert
  two-Higgs-doublet model with the monojet signal},''
  \href{http://dx.doi.org/10.1103/PhysRevD.99.015011}{{\em Phys.Rev.}
  {\bfseries D99} (2019) 015011},
  \href{http://arxiv.org/abs/1809.00933}{{\ttfamily arXiv:1809.00933
  [hep-ph]}}.

\bibitem{Belyaev:2016lok}
A.~Belyaev, G.~Cacciapaglia, I.~P. Ivanov, F.~Rojas-Abatte, and M.~Thomas,
  ``{Anatomy of the Inert Two Higgs Doublet Model in the light of the LHC and
  non-LHC Dark Matter Searches},''
  \href{http://dx.doi.org/10.1103/PhysRevD.97.035011}{{\em Phys.Rev.}
  {\bfseries D97} (2018) 035011},
  \href{http://arxiv.org/abs/1612.00511}{{\ttfamily arXiv:1612.00511
  [hep-ph]}}.

\end{thebibliography}
\providecommand{\href}[2]{#2}\begingroup\raggedright\endgroup

\end{document}